\documentclass[11pt]{article}
\pdfoutput=1
\usepackage[usenames,dvipsnames,svgnames,table]{xcolor}

\usepackage{booktabs}
\usepackage[english]{babel}
\usepackage{amsmath,amssymb,amsbsy,amstext, amsthm, simplewick}
\usepackage{hyperref}

\usepackage{tikz}
\usepackage{amsfonts}
\usepackage{amssymb}
\usepackage{upgreek}
\usepackage{simplewick}
 \usepackage{exscale,relsize}
\usepackage{mathtools}
\usepackage{comment}

\usepackage[margin=1cm,labelfont={sf,bf,scriptsize},textfont={sf,scriptsize}]{caption}

\usepackage{colortbl}
\definecolor{lightgreen}{cmyk}{0.2, 0, 0.2, 0.2}
\definecolor{lightgray}{cmyk}{0.1,0.2,0,0.1}
\definecolor{lightgray2}{cmyk}{0.1,0.1,0,0.1}

\makeatletter
\newlength{\apb@width}
\newcommand{\autoparbox}[2][c]{\settowidth{\apb@width}{#2}\parbox[#1]{\apb@width}{#2}}

\makeatother

\numberwithin{equation}{section}

\def\beq{\begin{equation}}
\def\eeq{\end{equation}}

\def\bea{\begin{eqnarray}}
\def\eea{\end{eqnarray}}

\def\beq{\begin{equation}}
\def\eeq{\end{equation}}
\def\be{\begin{equation}}
\def\ee{\end{equation}}
\def\bea{\begin{eqnarray}}
\def\eea{\end{eqnarray}}

\def\0{{\vec{0}}}

\DeclareRobustCommand{\SkipTocEntry}[4]{}

\def\beq{\begin{equation}}
\def\eeq{\end{equation}}

\def\ba#1\ea{\begin{align}#1\end{align}}
\def\bg#1\eg{\begin{gather}#1\end{gather}}
\newcommand{\bseq}{\begin{subequations}}
\newcommand{\eseq}{\end{subequations}}

\DeclareSymbolFont{extraup}{U}{zavm}{m}{n}
\DeclareMathSymbol{\varheart}{\mathalpha}{extraup}{86}
\DeclareMathSymbol{\vardiamond}{\mathalpha}{extraup}{87}

\newcommand{\es}[1]{\textcolor{blue}{(ES: #1)}}

\def\({\left(}
\def\){\right)}
\def\[{\left[}
\def\]{\right]}

\begin{document}

\begin{titlepage}
{~~~~~~~~~~~~~~~~~~~~~~~~~~~~~~~~~~~~
~~~~~~~~~~~~~~~~~~~~~~~~~~~~~~~~~~
~~~~~~~~~~~ \footnotesize{SLAC-PUB-16261,SU/ITP-15/05,NSF-KITP-15-046}} 
\setcounter{page}{1} \baselineskip=15.5pt \thispagestyle{empty}

\vbox{\baselineskip14pt
%\hbox{hep-th/0000000}
}
{~~~~~~~~~~~~~~~~~~~~~~~~~~~~~~~~~~~~
~~~~~~~~~~~~~~~~~~~~~~~~~~~~~~~~~~
~~~~~~~~~~~ }

\bigskip\

\vspace{2cm}
\begin{center}
{\fontsize{19}{36}\selectfont  \sc
String-theoretic breakdown of effective field theory near black hole horizons}
\end{center}

\vspace{0.6cm}

\begin{center}
{\fontsize{13}{30}\selectfont  Matthew Dodelson$^{1,2}$ and Eva Silverstein$^{1,3,4}$}
\end{center}

%\vspace{0.2cm}

\begin{center}
\vskip 8pt

\textsl{
\emph{$^1$Stanford Institute for Theoretical Physics, Stanford University, Stanford, CA 94306}}

\vskip 7pt
\textsl{ \emph{$^2$Kavli Institute for Theoretical Physics, University of California, Santa Barbara, CA 93106}}

\vskip 7pt
\textsl{ \emph{$^3$SLAC National Accelerator Laboratory, 2575 Sand Hill, Menlo Park, CA 94025}}

\vskip 7pt
\textsl{ \emph{$^4$Kavli Institute for Particle Astrophysics and Cosmology, Stanford, CA 94025}}

\end{center}

\vspace{1.2cm}
\hrule \vspace{0.3cm}
{ \noindent \textbf{Abstract} \\[0.2cm]
We investigate the validity of the equivalence principle near horizons in string theory, analyzing the breakdown of effective field theory caused by longitudinal string spreading effects. An experiment is set up where a detector is thrown into a black hole a long time after an early infalling string.  Light cone gauge calculations, taken at face value, indicate a detectable level of root-mean-square longitudinal spreading of the initial string 
as measured by the late infaller.  This results from the large relative boost between the string and detector in the near horizon region, which develops automatically despite their modest initial energies outside the black hole and the weak curvature in the geometry.  We subject this scenario to basic consistency checks, using these to obtain a relatively conservative criterion for its detectability.     In a companion paper, we exhibit longitudinal nonlocality in well-defined gauge-invariant S-matrix calculations, obtaining results consistent with the predicted spreading albeit not in a direct analogue of the black hole process.            
We discuss applications of this effect to the firewall paradox, and estimate the time and distance scales it predicts for new physics near black hole and cosmological horizons.    \vspace{0.3cm}
 \hrule

\vspace{0.6cm}}
\end{titlepage}

\tableofcontents
\section{Introduction}

In black hole physics, there is a large relative boost between early matter and late-infalling observers at the horizon.  This must be taken into account in estimating the breakdown of effective field theory (EFT) in string theory (or any UV completion of gravity), in order to reliably describe the interactions of a late infaller near the horizon, as explained recently in \cite{backdraft}\ in the context of a particular string production effect.\footnote{See e.g. \cite{lennyspreading}, \cite{JoeBHcomp}\ and  \cite{Giddingsboost}, among many others, for earlier discussions raising the question of the role of relative boosts and nonlocality in the black hole problem, albeit  focused more on information transfer than on horizon `drama'.}  In particular, the small curvature in Planck and string units does not {\it a priori} bound the level of  EFT violation in string theory:  the long time evolution can build up large effects even in a weakly curved background such as the exterior of a Schwarzschild black hole.       

Specifically, the relative boost in the near-horizon Rindler region between two time-translated but otherwise identical observers is given by a rapidity $\Delta\eta= \Delta t/2r_\text{s}$, where $\Delta t$ is the time translation in Schwarzschild coordinates, and $r_\text{s}$ is the Schwarzschild radius. 
Although the Schwarzschild geometry is weakly curved outside and near the horizon, the black hole functions as a powerful accelerator: 
given fixed Schwarzschild energy $E\sim m$ at which two observers of mass $m$ are dropped or thrown into the black hole, they reach the near horizon region with a large center of mass energy $E_{\text{CM, Rindler}}\sim m \exp(\Delta\eta/2)$ at late times.  In the Rindler region, the presence of such a large energy in the problem raises the important question of whether extended-string effects could be significant in describing the fate of the late infaller.  This question is especially urgent given existing evidence for enhanced string-theoretic non-adiabatic and spreading effects \cite{backdraft}\cite{lennyspreading} at large relative boost.  

As an approach to quantum gravity, string theory has passed many consistency checks, including providing a count of black hole microstates in special calculable examples \cite{StromingerVafa}.  It therefore seems reasonable to check carefully if it might contain the dynamics required to resolve other questions in black hole physics -- especially given the substantial theoretical uncertainty just noted in estimates of the breakdown of effective field theory in the regime $\Delta \eta \gg 1$.\footnote{There has been much interesting work on string theory scattering and black hole dynamics (see e.g. \cite{BHstringscatt}), with the ambitious goal of describing black hole formation in string theory (or obstructions to it).  Here we are concerned the string-theoretic interactions of a late probe sent into the black hole at a time $\Delta t\gg 2r_{\text{s}}$ after early string matter.  The latter may be thought of as a proxy for the matter that formed the black hole, but for our present purposes we can analyze the much simpler problem of early and late string probes of an existing black hole background.}   In this and a companion paper, we take several steps towards addressing this question, applying the approach of \cite{backdraft}\ to fundamental strings.  

This question is especially timely in light of recent work sharpening the paradoxes arising from black hole thought experiments \cite{firewalls}\ which involve a late-infalling observer.
%\footnote{There are many other important papers in the subject which sharpen the information problem, such as those discussing information transfer (REFs); we will focus on AMPS because of its emphasis of the thought experiments involving a late infaller.}  
The AMPS paradox \cite{firewalls}\ confirms the incompatibility of a complete quantum mechanical description and the approximation of low-energy EFT in black hole physics, a longstanding problem.     
We will present concrete evidence that string theory intrinsically  provides  `drama' for a late observer of the kind required to avoid more drastic resolutions of the AMPS paradox such as ad hoc modifications of gravity or violations of quantum mechanics.   
%There is much work in the literature on approaches modifying gravity and on approaches giving up quantum mechanics.   The latter is especially insane, although bizarrely popular.  
Since string theory provides a good candidate for a UV completion of gravity respecting ordinary quantum mechanics, it would be very satisfying, albeit somewhat less revolutionary, if it simply contains the required physics.  

In this paper we will show using simple kinematics that if the longitudinal string spreading obtained via worldsheet calculations in light-cone gauge \cite{lennyspreading}\ is accurate, then string theory provides interactions (hence violation of EFT) for the late infaller over a broad range of parameter space.  Along the way, we will refine our understanding of the longitudinal spreading prediction and its detectability, resolving various puzzles associated with the distinction between the longitudinal and transverse directions.    

Although we view string-theoretic drama as a relatively conservative prospect, we cannot regard it as the null hypothesis for the purposes of the present work aimed at assessing whether it is a viable possibility.    
In particular, we must address the validity of longitudinal spreading beyond the gauge-fixed calculation \cite{lennyspreading}, a subtle problem \cite{sixauthor}\cite{JoeBHcomp}.  
Using concrete scattering experiments in a companion paper, we will exclude the null hypothesis -- approximate local EFT for string probes of black holes -- at reasonable significance by exhibiting, among other things, a time advance which indicates non-local string interactions in the presence of large relative boosts.  This is contingent on the logarithmic behavior of the relatively well-established transverse string spreading.        

It is worth noting that the effects we find are not sharply localized at the horizon, since the large relative boost pertains once the probes reach the near-horizon regime $|r-r_{\text{s}}|\ll r_{\text{s}}$.  As such, it may be a concrete example of less-violent nonlocality than the specific proposal for a `firewall' in the AMPS paper \cite{firewalls}, although spreading-induced interactions between early string matter and (regular or mined) Hawking modes may introduce additional effects.  Being stringy, our effect does not correspond to \cite{NVNL}\ either.  There are two ways in which EFT might break down as delineated in the list of incompatible postulates described in \cite{firewalls}, depending on whether the effects extend outside the stretched horizon; on the face of it, our effect violates both. Because of this and the relatively short timescale in the problem, it is interesting to explore whether it may ultimately be testable similarly to the scenario explored in \cite{SteveObservations}.  
%using various methods of imaging black hole horizons (REFs).  \es{This should be stated extremely conservatively if we want to be taken seriously in astro...}  
We will also address the constraints from observational cosmology, where certain aspects of horizon physics are well tested.   We will conclude with a summary of our results and some important caveats as well as a discussion of future directions.  
     
\section{The size of strings}
\subsection{Longitudinal spreading in light-cone gauge}
Ideally, one would measure the size of strings (or any other extended object) using on-shell scattering experiments. In this section we will instead review how to compute the root mean square variation of the spacetime coordinates of the string in the worldsheet CFT \cite{lennyspreading,stringsize}. Although this calculation is both gauge-fixed and off-shell, it provides useful intuition for why strings are extended. We will work in the light-cone gauge on the worldsheet, which has the advantage of being a Hamiltonian quantum mechanical system with local interactions \cite{lightcone} (in contrast to the conformal gauge). 

It would be interesting to understand how to reproduce the same results in the conformal gauge, where there is no {\it a priori} separation of the target space coordinates into light cone and transverse directions.  Although we will not present detailed calculations in this covariant gauge, we will address the essential puzzle it raises below in Section \ref{puzzles}\ which will lead us to an
%and we will comment further on this below in order to motivate an 
identification of the light cone directions with the direction of relative motion between a string and a detector.  
It would also be interesting to generalize the calculations to static gauge, particularly since this would line up with `nice slice' coordinates in black hole backgrounds.\footnote{We thank J. Polchinski for discussions.}  However, the ultimate (thought-)experimental test comes from gauge-invariant observables, such as S-matrix elements.  In a companion paper \cite{usSmatrix}, we report on an analysis of four and five-point string amplitudes which exhibits concrete evidence for longitudinal nonlocality in the string S-matrix.   \\
\indent We define the light-cone coordinates as $x^{\pm}=x^0\pm x^1$. Choosing the light cone gauge $X^-=x^-+\alpha' p^-\tau$, the mode expansions of the transverse fields  of the closed string are
\begin{align}
X^i(\tau,\sigma)=x^i+\alpha' p^i\tau+i\left(\frac{\alpha'}{2}\right)^{1/2}\sum_{n\not=0}\frac{1}{n}\left(\alpha_n^i e^{-in(\tau-\sigma)}+\tilde{\alpha}_n^i e^{-in(\tau+\sigma)}\right)\label{transversemode}.
\end{align}
We can compute the average transverse deviation away from the center of mass of the string at time $\tau=0$,
\begin{align}
\langle (\Delta X^i)^2\rangle = \langle (X^i(0,\sigma)-x^i)^2\rangle=\alpha' \sum_{n>0}\frac{1}{n}.
\end{align}
This sum is formally infinite, but should be physically cut off at the maximum mode number $n_{\text{max}}$ that can be probed by the detector. The size then becomes
\begin{align}\label{transpread}
\langle (\Delta X^i)^2\rangle\sim \alpha' \log n_{\text{max}}.
\end{align}
In this expression we have assumed that $n_{\text{max}}$ is large, so that the detector has a very good time resolution.\\
\indent Now let us repeat the same analysis for the longitudinal coordinate $X^+$, which has the mode expansion 
\begin{align}
X^+(\tau,\sigma)=x^++\alpha' p^+\tau+i\left(\frac{\alpha'}{2}\right)^{1/2}\sum_{n\not=0}\frac{1}{n}\left(\alpha_n^+ e^{-in(\tau-\sigma)}+\tilde{\alpha}_n^+ e^{-in(\tau+\sigma)}\right)\label{longmode}.
\end{align}
Here the $\alpha_n^+$ oscillators can be expressed in terms of the $\alpha_n^i$ oscillators by solving the Virasoro constraints. They satisfy the Virasoro algebra,
\begin{align}\label{Virasoro}
[\alpha_m^+,\alpha_n^+]=\left(\frac{2}{\alpha'}\right)^{1/2}\frac{1}{p^-}(m-n)\alpha_{m+n}^++\frac{4m^3}{\alpha' (p^-)^2}\delta_{m+n}.
\end{align}
The size of the longitudinal coordinate is then
\begin{align}\label{longspread}
\langle (\Delta X^+)^2\rangle&\sim \frac{1}{(p^-)^2}\sum_{n>0}n\sim \frac{n_{\text{max}}^2}{(p^-)^2}.
\end{align}
Note that the correlation functions of $X^+$ do not satisfy a Gaussian distribution, in contrast to those of $X_\perp$. In order to understand the distribution of $X^+$ in more detail, one would need to compute the higher moments $\langle(\Delta X^+)^{2n}\rangle$. We were not able to do this calculation, but fortunately the root-mean-square size of the string is sufficient for our purposes. \\
\indent The constraint relating the transverse and longitudinal variables has helped obscure the role of the latter, since $X^+$ is not an independent variable.  It is worth emphasizing that there are many familiar situations in which such constraints apply, but the analogue of the longitudinal variables plays a crucial role in the physics.  Perhaps the most basic is the Hamiltonian constraint in gravity which relates the expansion of the universe to stress energy sources which are analogous to the transverse degrees of freedom here.  Although the scale factor is constrained, the expansion of the universe is obviously a physical effect.  On the string worldsheet, similarly the expansion and contraction of the string in the direction of its extent are physical (as in the `yo yo' solutions we review in \cite{usSmatrix}).  

\indent It is instructive to repeat the calculation for open strings. In this case one finds
\begin{align}
\langle (\Delta X^i)^2\rangle&\sim\alpha' \sum_{n>0}\frac{1}{n}\cos^2(n\sigma)\\
\langle (\Delta X^+)^2\rangle&\sim \frac{1 }{(p^-)^2}\sum_{n>0}n\cos^2(n\sigma).
\end{align}
At the endpoints of the string the results are the same as for closed strings. On the other hand, the midpoint of the string at $\sigma=\pi/2$ does not fluctuate at all. 

The decomposition into $X^\pm$ and the transverse directions obscures the Lorentz invariance of the theory.   This raises the question of whether the difference in length scales obtained above for longitudinal (\ref{longspread}) versus transverse (\ref{transpread}) string spreading is consistent with Lorentz invariance.   In a $2\to n$ scattering process, the relative motion of the two incoming legs picks out a preferred direction.  The difference in length scales found above for longitudinal and transverse spreading is consistent with Lorentz invariance if the longitudinal direction is identified roughly with the direction of relative motion between the string and detector.  After developing the notion of the detector and its resolution further in the next section, we will attempt to make this identification precise.

As the relative boost between the source string and the detector increases, the detector resolution improves, giving it sensitivity to more of the modes contributing to the spreading 
(\ref{transpread}-\ref{longspread}).  This improved sensitivity follows from the time dilation between the detector and source clocks. 
In the context of parton physics in \cite{lennykogut}, the limiting case of this is known as the infinite momentum frame, where one boosts the hadron to infinite momentum so that its internal motions are arbitrarily slow due to the time dilation.

\subsection{Strings as detectors}
Now that we have derived the size of strings in terms of the detector parameter $n_{\text{max}}$, we specialize to the case where the detector is itself is a string. We now will determine the shortest time scale that a string can probe, which is the inverse of the maximum frequency $n_{\text{max}}$ in units of $\alpha'$.\\
\indent Let us assume that the detector is a string with momentum $p_2$, and that the string being probed has momentum $p_1$. Under time evolution with the light-cone Hamiltonian, the state of String 2 evolves as
\begin{align}
|p_2^\mu,X^-\rangle=e^{-ip_2^+X^-}|p_2^\mu\rangle
\end{align}
The natural oscillation time on the detector is then 
\begin{align}\label{deltax-}
\Delta X^-\sim \frac{1}{p_{2}^+}.
\end{align}
In terms of the light-cone time on the string being measured, $\tau= X^-/(\alpha' p_1^-)$, this gives an oscillation time
\begin{align}
\Delta \tau\sim \frac{1}{ \alpha' p_2^+p_1^-}\label{deltatau}.
\end{align}
This oscillation time acts as an internal clock, which cannot probe times that are far shorter than the scale (\ref{deltatau}). It follows that the largest mode number the detector could potentially probe is
\begin{align}
n_{\text{max}}&\lesssim \frac{1 }{\Delta \tau}\sim \alpha'p_2^+p_1^-\label{nmax}.
\end{align} 
As we will see shortly, the frequency cutoff of a given scattering process can be smaller than (\ref{nmax}), so (\ref{nmax}) represents an upper bound on $n_{\text{max}}$.\\
\indent  For $p_2^-\not=0$, it is instructive to write (\ref{nmax}) as 
\begin{align}
n_{\text{max}}\lesssim \alpha' \frac{p_1^-}{p_{2}^-}({p}_{2\perp}^2+m^2).
\end{align}
Note that this mode number increases with the relative longitudinal boost between the two strings. This is a consequence of the fact that the time dilation slows down the internal motion of String 1, which gives the detector a finer time resolution as the boost increases.  

\indent Let us now compare the frequency cutoff (\ref{nmax}) to the maximum mode number probed in the Regge limit $s\gg t$ of string scattering amplitudes, as analyzed in \cite{bpst}.  Specifically, we will quote the results of Section 4 of \cite{bpst}, translated to our notation and conventions.   Consider the $s$-channel diagram for $2\to 2$ scattering $(p_1,p_2)\to (p_3,p_4)$ in light-cone gauge, in a ``brick wall'' frame where $p_1^+\approx 0$, $p_1^+=p_3^+$, and $p_2^+=p_4^+$.  We also restrict to external states of negligible mass for now, although we will briefly discuss the role of the mass below. This diagram can be expressed as an integral over the time $T=\Delta X^-$ between the joining and splitting interaction vertices, with integration measure
\begin{align}
d\mu(T)=\frac{1}{T^2}e^{p_2^+T}.
\end{align}
This measure factor is integrated against the contribution from the oscillators, after performing the Gaussian integral over the modes $X_{\perp n}$ of the transverse embedding coordinates of the string.  The latter explicitly produces a factor
\be\label{moneyfactor}
\exp\left(-\sum_n \frac{k_\perp^2\alpha'}{n + T n^2/(2\alpha' p_1^-)}\right)
\ee
where $|k_\perp| = 2 |p_{\perp, r}|, r=1,\dots, 4$ is the absolute value of the momentum transfer. See equations (4.19)-(4.20) of \cite{bpst} for more details (note that we have different conventions for the light cone coordinates and the incoming and outgoing string labels). The maximum mode number that contributes to the oscillator sum on String 1 is then
\begin{align}
n_{\text{max}} = \frac{2\alpha' p_1^-}{|T|}.
\end{align}
This is very precise, since
\be\label{sumsagree}
\sum_{n=1}^{n_{\text{max}}}\frac{1}{n} = \sum_{n=1}^\infty \frac{1}{n+n^2/n_{\text{max}}}.
\ee
For $s\gg-\alpha' t\gg1$ the integral over $T$, defined by analytic continuation as described in \cite{bpst},  is dominated at a saddle point,
\begin{align}\label{Tsaddle}
T\sim - \frac{\alpha' p_\perp^2}{p_2^+}.
\end{align}on
It follows that the frequency cutoff is at
\begin{align}
n_{\text{max}}\sim \frac{p_1^-p_2^+}{p_\perp^2}\label{nmaxcons}.
\end{align}
Since $p_\perp^2\gg 1/\alpha'$ in this saddle point calculation, the mode cutoff is bounded above by the right hand side of $(\ref{nmax})$.\\ 
\indent As the momentum transfer $k_\perp^2$ approaches zero, the validity of the saddle point (\ref{Tsaddle}) is lost.   It would be very interesting to understand the detectability of longitudinal string spreading in that regime.  

\indent Using the results of the previous section, we find that if 
\be\label{nmaxlarge}
n_{\text{max}}\sim \frac{ p_{2}^+p_1^-}{p_\perp^2} \gg 1,
\ee
 then the size of String 1 as seen by String 2 is \begin{align}
\langle (\Delta X^i)^2\rangle&\sim \alpha' \log (s/t)\label{transverse}\\
\langle (\Delta X^+)^2\rangle&\sim \left(\frac{ p_{2}^+}{p_\perp^2}\right)^2\label{deltax+}.
\end{align}
In the first expression we used that $p_\perp$ is of order the momentum transfer in the brick wall frame defined in \cite{bpst}. We again emphasize that the computation has been done for negligible string mass.   
%More precisely, this is the size of String 1 assuming that String 2 is an ideal detector, so it is an upper bound on the size of String 1. 
Note that these expressions have the correct transformation laws under the subgroup of Lorentz transformations that are preserved by the light-cone gauge.  

It would be interesting to generalize this computation of \cite{bpst}\ to higher oscillator levels of the external strings, to capture the case of massive detectors or source strings.    Folding in Hermite polynomials as a function of the modes of the transverse embedding coordinates $X_\perp$ introduces nontrivial mass dependence, which according to our preliminary calculations can introduce a term $\propto m^2$ added to the $p_\perp^2$ in the above expressions, although we have not established whether or not this is general.\footnote{In the brick wall frame, each transverse momentum $p_{\perp r}$ of the external strings carries half of the momentum transfer $k_\perp$.  Partly because of this ambiguity between the dependence on momentum transfer and on the transverse momentum carried by the external states, it is not immediately clear whether the result must be to replace $p_\perp^2$ with the relativistic combination $p_\perp^2+m_2^2$, with $m_2$ the mass of the detector string (although below in Section \ref{puzzles}\ we will present a somewhat different argument that this provides a conservative estimate for the spreading).}    
As we will see, the case of massive detectors will be of interest in the black hole problem.  To make use of existing calculations, we can consider as an example the particular situation where the string mass comes from transverse Kaluza-Klein momentum in some extra dimension, and use the results just presented with the Kaluza-Klein momentum playing the role of $p_\perp$.  The generalization to oscillators could also enhance the effect in a different way, via the Hagedorn density of available final string states which played a role in \cite{backdraft}.  We will not need that to establish an interesting breakdown of EFT in the present work, but it would be interesting to incorporate in a more general analysis of scattering in the massive case.

\indent The condition (\ref{nmaxlarge}) is necessary to justify the approximations that we have made in the mode sums (\ref{transpread}) and (\ref{longspread}).    This inequality involves the source string 1 and can be satisfied by increasing the relative boost between the strings.  The formula (\ref{deltax+}) for the detectable longitudinal spreading also has an interesting structure.  In Section \ref{puzzles}\ below, we will argue based on basic features of the geometry of the trajectories that there is an intuitive reason that the detected 
spreading is not in fact determined solely by $p_2^+ \sim 1/\Delta X^-$:  instead, this putative spreading is only actually measurable by the detector string if the longitudinal direction is chosen to be the direction of relative motion between the source and detector.  The brick wall frame defined in \cite{bpst}\ gives the only notion of longitudinal direction which is invariant under time reversal symmetry, as we describe below around equation (\ref{longBW}).\\ 
\indent It is interesting to consider the implications of (\ref{deltax+}) in different frames. First suppose that the energy of the detector is held fixed as the energy of the string being probed increases, as would be the case in the lab frame. Then the maximum detectable longitudinal size (\ref{deltax+}) remains constant with the energy of String 1. Said differently, the string being measured fails to Lorentz contract as emphasized in \cite{lennyspreading}. On the other hand, if we work in the center of mass frame, then the size of String 1 grows linearly with the energy, $\Delta X^+\sim \alpha' E/k_\perp^2$.\footnote{The nonlocality of scattering in string theory has previously been analyzed in \cite{grossmende} in the context of the hard scattering limit. In that work it was pointed out that the worldsheet saddle point for the string scattering amplitude in conformal gauge takes the form
%\begin{align}
$X^\mu(z,\overline{z})=i\sum_i k_i^\mu\log(z-z_i)\label{saddle}.$
%\end{align}
This scales linearly with the longitudinal momenta of the strings, which resembles the result (\ref{deltax+}). A longitudinal interaction region that grows linearly with energy has also been proposed in the context of hadronic physics \cite{lennykogut,gribov}.}\\
\indent Combining (\ref{deltax-}) and (\ref{deltax+}), and for $t\sim -p_\perp^2$ of order $1/\alpha'$, one finds\footnote{We thank S. Giddings for suggesting this.} 
\begin{align}\label{sup}
\Delta X^+\, \Delta X^-\sim \alpha'.
\end{align}
This is reminiscent of the stringy uncertainty principle $\Delta T\,\Delta X\ge \alpha' $ \cite{sup}.
% due to its resemblance to the Heisenberg uncertainty principle with $\hbar$ replaced with $\alpha'$.

\subsection{Longitudinal versus transverse directions}\label{puzzles}

The computations \cite{lennyspreading}, as reviewed above, derived the maximum potentially detectable longitudinal spreading as $\alpha' p_{2}^+$, but did not establish under what circumstance String 2 actually interacts with String 1 at that distance with an appreciable amplitude.  The calculation of \cite{bpst}\ just reviewed established the cutoff (\ref{nmaxcons}) for the case where the masses are negligible and the detection proceeds via $2\to 2$ massless string scattering, revealing a suppression factor of $1/\alpha' p_\perp^2$.  If the formula (\ref{deltax+}) did not depend on $p_\perp$ in the brick wall frame it would have been puzzling, since without that factor the measured spreading would depend only on the motion of the detector string (although the source string comes into the requirement (\ref{nmaxlarge})).  

Taking a step back, having light cone time resolution of at least $\Delta X^-\sim 1/p_{2}^+$ is a necessary condition that String 2 must satisfy in order to measure a spreading of the source string 1 of order $\Delta X^+\sim \alpha'/\Delta X^-$.  Whether it does so may depend on additional factors, such as $p_\perp$ in the case of $2\to 2$ ultrarelativistic string scattering just discussed.   On general grounds, the answer to this question must respect the underlying Lorentz invariance of the theory, reconciling this symmetry with the distinct scales of longitudinal and transverse spreading suggested by (\ref{transverse})-(\ref{deltax+}). 

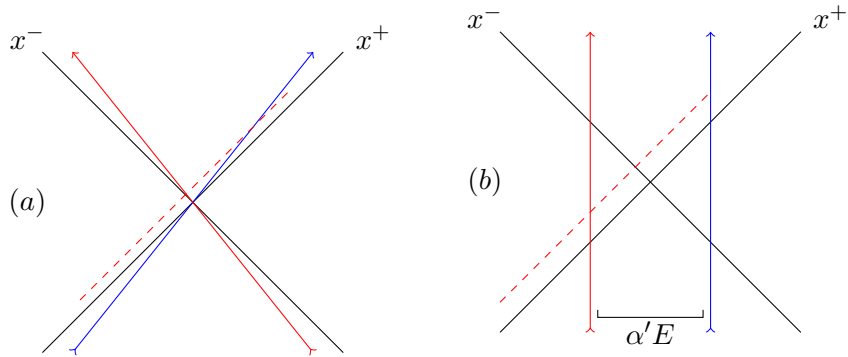
\begin{figure}
\begin{center} \begin{tikzpicture}[scale=2]
\draw (-1.1,0) node{$(a)$};
 \draw (-1,-1) -- (1,1);
 \draw (1,-1) -- (-1,1);
 \draw[>->,color=blue] (-.8,-1) -- (.8,1);
 \draw[>->, color=red] (.8,-1) -- (-.8,1);
 \draw[dashed, color=red] (-.75,-.65) -- (.65,.75);
  \draw (1.2,1.1) node{$x^+$};
    \draw (-1.1,1.1) node{$x^-$};
 \end{tikzpicture}
 \hspace{5 mm}
\begin{tikzpicture}[scale=2]
\draw (-1.1,0) node{$(b)$};
 \draw (-1,-1) -- (1,1);
 \draw (1,-1) -- (-1,1);
 \draw[>->,color=red] (-.4,-1) -- (-.4,1);
 \draw[>->, color=blue] (.4,-1) -- (.4,1);
 \draw[dashed, color=red] (-1,-.8) -- (.4,.6);
  \draw (1.2,1.1) node{$x^+$};
  \draw (-.35,-.85) --(-.35,-.9) -- (.35,-.9) -- (.35,-.85);
  \draw (0,-1) node{$\alpha' E$};
    \draw (-1.1,1.1) node{$x^-$};
 \end{tikzpicture}
\end{center}
    \caption{ Two configurations in which we assess the detectability of the spreading of the red string by the blue detector, as described in the text.  In the left panel (a), the direction of relative motion is along the prescribed light cone directions $x^\pm$.  In the right panel (b), the direction of relative motion is transverse to $x^\pm$.\label{figab}  }
\end{figure}
This question ultimately requires a gauge-invariant treatment, something we analyze explicitly using S-matrix amplitudes in \cite{usSmatrix}.  However, we can understand the essential features physically within the scope of our present analysis, as follows.
%Our expression (\ref{nmax}) for $n_{\text{max}}$ above raises the following puzzle at first glance.  
Consider the two configurations illustrated in Figure \ref{figab}.  These raise the following puzzle.  

Configuration (a) describes a pair of strings in relative motion along the $x^\pm=T\pm x$ directions, with energy $E\gg m$.  Configuration (b) describes a pair of strings with zero relative velocity in the $x^\pm$ directions, with energy $E=\sqrt{p_y^2+m^2}$ coming from a combination of transverse momentum and mass.  These strings are separated by a distance $ \alpha' E$ in $x$.

Before incorporating any details of the interaction between the source string and the detector, the light cone time resolution of String 2 is $\Delta X^-\sim 1/p_{2}^+\sim 1/E$ in both cases.  This in itself would endow String 2 with the necessary resolution to detect String 1, with the $p_\perp^2$ factor in the denominator in (\ref{deltax+}) replaced with $1/\alpha'$.  But in case (b), the strings are moving with large center of mass energy in the transverse $y$ direction.  If we had taken the light cone directions to be $y^\pm =T\pm y$, their large separation in $x$ would be in a transverse direction, in which they only spread logarithmically according to (\ref{transverse}).  As such, from that point of view one would not expect a significant interaction between the two strings.   

It is clear that in order to resolve this puzzle, there must be a way to uniquely define the longitudinal and transverse directions for a given scattering process.  This is not inconsistent with the spacetime symmetries since the direction of relative motion of the source string and detector breaks the rotational symmetry.   As mentioned above, the ``brick wall'' frame where $p_1^-\approx 0$, $p_1^+=p_3^+$, and $p_2^+=p_4^+$ is a particularly natural choice for computations in light-cone gauge \cite{bpst}. In this frame the worldsheet lengths of the incoming strings are equal to the lengths of the outgoing strings, and the scattering amplitude reduces to a quantum mechanical expectation value in the ground state of the string, with a frequency cutoff at (\ref{nmaxcons}) for mass $m=0$. This gives an operational meaning to the above computation of $\langle (\Delta X_\perp)^2\rangle$ and $\langle (\Delta X^+)^2\rangle$. \\
\indent The puzzle is therefore plausibly resolved by choosing the longitudinal direction so that $p_1^+=p_3^+$ and $p_2^+=p_4^+$. For example, for a scattering process where the initial particles are traveling in the $\pm x$ direction in the center of mass frame and scatter by an angle $\theta$, the longitudinal direction is 
\begin{align}\label{longBW}
x^+=T+x\cos(\theta/2)+y\sin(\theta/2).
\end{align}
For small angle scattering, this is roughly aligned with the relative direction of motion of the incoming states. Note that $x\cos(\theta/2)+y\sin(\theta/2)$ is one of two directions in the plane of scattering that is invariant under time reversal, the other being the transverse direction $y\cos(\theta/2)+x\sin(\theta/2)$. With this taken into account, the prediction of distinct scales for longitudinal and transverse spreading is internally consistent.  

So far, we have not established whether the spreading (\ref{transverse}) and (\ref{deltax+}) is detected by any observer with the correct time resolution if the longitudinal direction is chosen as above, although we have established this for massless strings using \cite{bpst}.  More generally,
there may be additional obstructions to the measurement. For example, let us take the detector to be a massive string moving in the $x$ direction, performing a measurement on a string moving in the $-x$ direction at scattering angle $\theta=0$. The trajectory of the detector satisfies
\be\label{atraj}
 \frac{dx^+}{dx^-} = \frac{1+v}{1-v}= \frac{(p_{2}^+)^2} {m^2},
\ee
where we used $p^+=\gamma m(1+v)$ with $v=dx/dT$.
It follows that in a resolution time $\Delta X^-\sim 1/p_{2}^+$, the detector string propagates along the light-cone space direction $x^+$ a distance of order $1/(\alpha' m^2)$ times the maximal detectable spreading radius $\alpha' p_{2}^+$.  More generally, at nonzero scattering angle we will have nonzero momentum transfer $p_\perp$, and the denominator in (\ref{atraj}) then contains a term $\sim p_\perp^2$.   

For large $m^2+p_\perp^2\gg 1/\alpha'$, the detector therefore only traverses a small fraction of the spreading region in a resolution time. This should be contrasted with detectors with $p_\perp^2+m^2\ll 1/\alpha'$, which propagate through the entire spreading region during a resolution time. We have not performed a complete analysis of the mass dependence of the cutoff $n_{\text{max}}$ using the S-matrix, although one can estimate this by folding Hermite polynomials into the calculation \cite{bpst}\ as discussed above.  In any case, it seems possible that the difference in trajectories implies that highly massive detectors can see less of the spreading region.\footnote{This may be related to an interesting question raised by Don Marolf about the interrogation time, which is the time over which the experiment occurs.} One possible modification of the result (\ref{deltax+}) that would take this difference into account is if a detector could only measure the portion of the spreading radius that it traverses in a resolution time. For a detector with mass $m^2>1/\alpha'$, this would imply that
\begin{align}\label{conservative}
\langle (\Delta X^+)^2\rangle=\frac{(p_2^+)^2}{(p_{2\perp}^2+m^2)^2}.
\end{align}
We regard (\ref{conservative}) as a conservative estimate of the spreading seen by a massive string. 
When we apply this to the black hole problem, we will find that it still allows a substantial breakdown of effective field theory for a late-infalling detector.  

%Later we will compare its implications for black holes to the more standard formula (\ref{deltax+}).

% \end{comment}

\subsection{Comparison to scattering amplitudes:  summary of \cite{usSmatrix}}

As already mentioned, although the spreading mechanism introduced in \cite{lennyspreading}\ and refined above is physically appealing, it is very important to determine whether the longitudinal RMS spreading has a measurable effect on gauge-invariant quantities.  Reconstructing the detailed finite-time dynamics within an S-matrix amplitude is not in general possible, but one can test for the effect in various ways.  For example, one can try to formulate precise tests for early interaction that could be induced by longitudinal spreading, taking as given the logarithmic transverse spreading in the $X_\perp$ directions (which we describe in detail to the required precision in \cite{usSmatrix}).   

In \cite{usSmatrix}\ we take this approach, analyzing $2\to 2$ and $2\to 3$ scattering.  Working in a Regge (or double Regge) regime at four (and five) points, we derive the phases in the amplitudes.  With the inclusion of wavepackets to localize the incoming and outgoing strings,  the phases determine their peak trajectories, including the peak impact parameter and time delays or advances.  

\begin{figure}
\begin{center}
\begin{tikzpicture}[scale=1.5]
\draw[>->,color=red] (-1.2,.5) -- (.31,.5) -- (1,-1);
  \draw[>->,color=blue] (1.2,-.5) -- (-.31,-.5) -- (-1,1);
  \draw[dashed] (-.31,-.5) -- (-.31,.5);
    \draw[dashed] (-.31,-.5) -- (.59,-.08);
        \draw[dashed] (-.31,-.5) -- (-1,-.5);
 \draw (.3,-.33) node {$b$};
  \draw (-.23,0) node {$b$};
    \draw (-1.5,0) node {$(a)$};
    \draw (-1.1,.35) node {$k_1$};
        \draw (1.1,-.35) node {$k_2$};
        \draw (.8,-.9) node {$k_3$};
                \draw (-.8,.9) node {$k_4$};
    \draw (-.5,-.4) node {$\theta$};

\end{tikzpicture}\hspace{20 mm}
 \begin{tikzpicture}[scale=1.5]
\draw (-1.5,0) node {$(b)$};
\draw(0,-1) -- (0,1);
\draw (-1,0) -- (1,0);
\draw[->,color=blue] (-.55,-.45) -- (.75,.85); 
\draw (.9,.9) node{$k_3$};
\draw[>->,color=red] (.7,-1) -- (.25,.8);
\draw[>->,color=red] (-.7,-1) -- (-.25,.8);
\draw (1,-.1) node{$x_1$}; 
\draw (-.1,1) node{$T$}; 
\draw (-.8,-.8) node{$k_1$}; 
\draw (.8,-.8) node{$k_2$}; 
\end{tikzpicture}
\end{center}
\caption{Two illustrative processes from the paper \cite{usSmatrix}.  They exhibit properties which fit with longitudinal spreading and are difficult to  interpret purely in terms of transverse string spreading as described in the text.\label{smatrixpic}}
\end{figure}
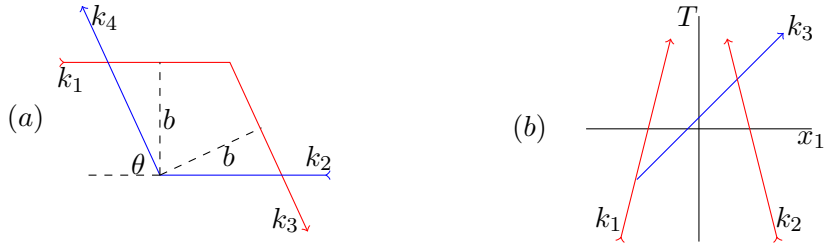

At four points, one of the open string orderings exhibits a nonzero peak impact parameter $b$, as indicated in Figure \ref{smatrixpic}a.  This fits with an explicit string `yo yo' solution describing an intermediate string created in the process, which reproduces the peak impact parameter and time delay.  
Tracing back the peak trajectories, they meet nontrivially in all dimensions, with the intermediate string solution fitting precisely into the resulting rhombus.   Further, at five points we find Bremsstrahlung radiation which traces back to the turning points in this process, providing a check on the simple picture resulting from the meeting of the traced-back trajectories.  Altogether, in an overconstrained problem, a very simple (a priori naive) picture of the intermediate scattering process fits the facts in a highly nontrivial way.

This structure does not jibe with a purely transverse process of string joining and splitting, assuming that works as in \cite{lennyspreading}\cite{bpst}, since the joining at nonzero impact parameter would be suppressed.  Early joining induced by longitudinal spreading does fit nontrivially with the structure of the amplitude, suggesting that it does play a role.   

In the diagram just described, the turning of each incoming string into an outgoing one occurs after the putative center of mass collision; there is a net time delay.  However,
another diagram at five points -- a perturbation of a diagram which at four points has zero time delay or advance -- exhibits advanced emission of one of the outgoing strings (this is depicted in Figure \ref{smatrixpic}b).  This indicates an early interaction, requiring longitudinal nonlocality.  Again, this statement is predicated on the limited transverse interaction derived in \cite{lennyspreading}\cite{bpst}.
%  and exhibit via explicit calculation a smoking-gun signature of longitudinal nonlocality, in the form of a time advance in five-point string scattering.  
%In addition, we show the even at the level of four-point amplitudes, the distribution of amplitudes in impact 
%parameter space lines up with expectations from longitudinal string spreading, 
For these reasons, derived in extensive detail in \cite{usSmatrix}, we conclude that longitudinal string spreading provides the simplest explanation that fits the `data' obtained from a wavepacket analysis of string amplitudes (including their phases).    

%The sharp result at five points and this more general picture strongly support the reality of longitudinal string spreading; it would be interesting to trace the S-matrix result back to the light cone analysis more directly.  

%Let us summarize those results briefly here; the interested reader may consult \cite{usSmatrix}\ for a step by step analysis.  

%\begin{figure}
%\begin{center}
%\begin{tikzpicture}[scale=1.5]
% \draw[>->,blue] (-1,-1) -- (1,1);
%  \draw (-1.1,-1.1) node {$k_1$};
% \draw(1.15,-1.15) node {$k_2$};
% \draw (1.1,1.1) node {$k_3$};
% \draw[>->,red] (1,-1) -- (-1,1);
% \draw(-1.15,1.15) node {$k_4$};
% \draw[gray,very thin] (0,0) -- (.2,.4) -- (-.2,.2) -- (0,.6) -- (-.4,.4) -- (-.2,.8) -- (-.6,.6) -- (-.4,1) -- (-.8,.8) -- (-.6,1.2) -- (-1,1);
% \end{tikzpicture}
% \end{center}
%  \caption{A logical possibility is that the spreading is triggered by the interaction of the center of masses. \label{weak}}
% \end{figure}

\section{Interactions of strings near black hole horizons}
\subsection{Notation and kinematics}
We now turn from flat space string theory to string theory on a black hole background. The metric takes the standard form, depicted in Figure \ref{coordinates},
\begin{align}
ds^2=-\left(1-\frac{r_{\text{s}}}{r}\right)\, dt^2+\left(1-\frac{r_{\text{s}}}{r}\right)^{-1}\, dr^2.
\end{align}
Here $r_{\text{s}}=2G_{\text{N}}M_{\text{BH}}$ is the Schwarzschild radius of the black hole and we have suppressed the transverse coordinates.   

The near horizon region $|r-r_{\text{s}}|\ll r_{\text{s}}$ is approximately flat, reducing to a portion of Minkowski spacetime; in this region the geometry is nearly identical to the near-horizon region of Rindler space.  We will be interested in the behavior of early and late-infalling trajectories when they reach this region.  

\begin{figure}
\begin{center}
\begin{tikzpicture}[scale=2.5]
 \draw[->] (0,0) -- (1,1);
 \draw (1.1,1.1) node {$x^+$};
 \draw[->] (1,-1) -- (-1,1);
 \draw(-1.1,1.1) node {$x^-$};
 \draw(-1,1) .. controls (0,.75) .. (1,1);
 \draw[dotted](1,-1) .. controls (.25,0) .. (1,1);
 \draw(.9,0) node {\small$r=\text{constant}$};
 \draw(.25,.5) node {\small$r=r_{\text{s}}$};
  \end{tikzpicture}
  \end{center}
\caption{\label{coordinates}}
 \end{figure}

It will be useful to use Kruskal coordinates, 
\begin{align}
ds^2=-\frac{2r_\text{s}}{r}e^{1-r/r_{\text{s}}}dx^+\, dx^-.
\end{align}
obtained for $r>r_{\text{s}}$ from the transformation 
\begin{align}
x^{\pm}= \pm \sqrt{2r_\text{s}(r-r_{\text{s}})}\exp\left(\frac{r-r_{\text{s}}\pm t}{2r_\text{s}}\right)\label{transform}.
\end{align}
This makes manifest that a translation in Schwarzschild time acts as a Lorentz boost in the near horizon region,
\begin{align}
x^{\pm}(t+\Delta t)=\exp\left(\pm \frac{\Delta t}{2r_{\text{s}}}\right)x^{\pm }(t)\label{boost}.
\end{align}
This boost becomes large for long time separations $\Delta t\gg r_{\text{s}}$. \\
\indent The setup of our thought experiment consists of an observer stationed at a constant Schwarzschild radius $R\gg r_{\text{s}}$.  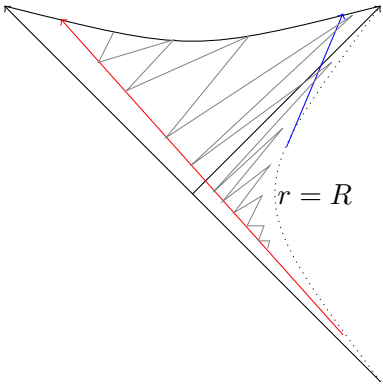
\begin{figure}
\begin{center}
\begin{tikzpicture}[scale=2.5]
 \draw[->] (0,0) -- (1,1);
 \draw[->] (1,-1) -- (-1,1);
 \draw(-1,1) .. controls (0,.75) .. (1,1);
 \draw[dotted](1,-1) .. controls (.25,0) .. (1,1);
 \draw[->,red] (.8,-.75) -- (-.7,.93);
 \draw[->,blue] (.5,.25) -- (.8,.96);
 \draw[gray,very thin] (.4,-.29) -- (.41,-.25) -- (.35,-.25) -- (.38, -.17) -- (.29,-.17) -- (.37,-.01) -- (.22,-.1) -- (.41,.15) -- (.16,-.04) -- (.48,.35) -- (.11,.01) -- (.74,.7) -- (-.01,.15) -- (.85,.95) -- (-.14,.3) -- (.3,.84) -- (-.35,.55) -- (-.1,.82) -- (-.5,.70) -- (-.42,.86);
 \draw(.65,0) node {$r=R$};
  \end{tikzpicture}
  \end{center}
  \caption{A cartoon of the proposed thought experiment. The early and late strings are displayed in red and blue respectively. The zigzag lines indicate the spreading of the early string as detectable by the late string, which only develops sufficient resolution to see it in the near horizon region.  The dotted line is a constant $r$ locus from which both detector and string may be dropped at different times.}
 \end{figure}
 The observer sends in two strings, separated by a long time interval $\Delta t\gg r_{\text{s}}$. Because of (\ref{boost}), the center of mass energy of the two strings is very large in the near horizon region, so the nonlocal effects of the previous section may become important. On the other hand, the strings are separated by a significant interval $x^+_2-x^+_1$ along the horizon, which we should compare to the detectable spreading radius.

 In our analysis, we will assume that when String 1 reaches the near horizon region, it has evolved into the standard single-string state in flat spacetime.  Given that, in the near horizon region we can use the refined calculation of \cite{lennyspreading}\ explained in the previous section.
 As developed above in (\ref{deltax+}) and Section \ref{puzzles}, under these conditions, the criterion for second string to feel the presence of the first is 
\begin{align}
x^+_{21,\text{h}}< \frac{{p^+_{2,\text{h}}}}{p_{\perp}^2+m_2^2}.
\end{align}
where $x^+_{21,\text{h}}=x^+_{2,\text{h}}-x^+_{1,\text{h}}$ is the coordinate separation along the null direction $x^+$ between Strings 1 and 2 as they cross the horizon, and $p_\perp$ is the momentum transfer.   A subscript `h' signifies that the quantity should be evaluated at the horizon.  It is worth noting that this is not a strict inequality, in that one can relax this condition by an order one factor on the right hand side, although the 
%Gaussian 
falloff at large $X^+$ in the distribution 
%(\ref{Virasoro}) 
of the nominal spreading (\ref{longspread}) would induce a corresponding penalty in amplitude.\footnote{We have not computed this distribution precisely.   Because the constraint relating the longitudinal embedding coordinate to the Gaussian-distributed transverse modes is $\partial_{\sigma_\pm} X^+\sim (\partial_{\sigma_\pm}X_\perp)^2$, we suspect that at large $X^+$ the distribution may be linear in the exponent, i.e. of the form $\exp(-|X^+|/\sqrt{\langle(\Delta X^+)^2\rangle})$ for sufficiently large $\Delta X^+$.} 

As described above, the saddle point analysis leading to (\ref{Tsaddle}) requires $p_\perp^2 > 1/\alpha'$, or a similar scale of transverse Kaluza-Klein momentum which enters our formulas as a mass term.   We do not have an explicit understanding of the cutoff $n_{\text{max}}$ away from this regime.  We will therefore mostly have in mind masses and/or transverse momenta above the string scale, but its precise value will not play an important role in our basic results. The mass $m_2$ here could be made up of Kaluza-Klein momentum from extra dimensions, or possibly oscillator contributions to the string mass, following the conservative criterion (\ref{conservative}).    \\
%For the more conservative estimate (\ref{conservative}), 
%one should replace $\alpha'$ in this formula with $1/m_2^2$.
\indent More elaborate scenarios in which String 2 emits a third string 3 in the near-horizon region lead to a similar condition for detecting a violation of effective field theory,
\be\label{ellcondgen}
x^+_{31,\text{h}} <  \frac{{p^+_{3,\text{h}}}}{p_{\perp}^2+m_3^2}.
\ee
This generalization will be important below.  
%Here we used (\ref{deltax+}) and the fact that the light cone directions $x^\pm$ in our near-horizon flat spacetime region are in the plane of relative motion of the early and late string, as required by the criterion developed in \S\ref{puzzles}.  That is, the longitudinal string spreading $\sim \alpha' p_{2-}$ applies along the $x^\pm$ directions since these correspond to the direction of relative motion of the two strings.   In the next two sections, we will show that these conditions are satisfied for an interesting range of controlled parameters. 

\begin{figure}
\begin{center}
\begin{tikzpicture}[scale=2.5]
 \draw[->] (0,0) -- (1,1);
 \draw[->] (1,-1) -- (-1,1);
 \draw(-1,1) .. controls (0,.75) .. (1,1);
 \draw[dotted](1,-1) .. controls (.25,0) .. (1,1);
  \fill[color=red]  (.8,-.75) -- (.64,-.5) --  (-.6,.9) -- (-.7,.93);
    \fill[color=blue]  (.53,.3) -- (.6,.4) --  (.67,.93)-- (.6,.9);
  \end{tikzpicture}
\end{center}
\caption{Proportions in our setup.  The separation between 1 and 2 is much larger than the intersection of String 1's worldsheet with the horizon, our condition (\ref{separate}). \label{figbhws}  
}
\end{figure}
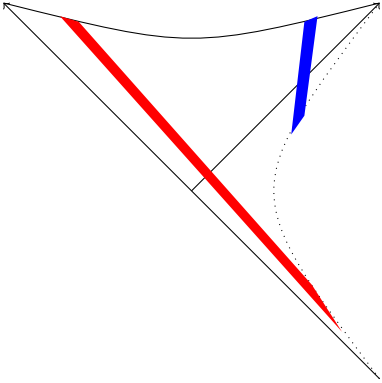

In order for the interaction to be nontrivially generated by the longitudinal spreading -- as opposed to an effect that is basically contained within effective field theory -- we need the separation $x_{21,\text{h}}^+$ to be large enough that the two worldsheets do not directly intersect according to their fiducial sizes.  The fiducial proper size of a string is of order the string length $\sqrt{\alpha'}$ for a string-scale mass,  of order $(m\sqrt{\alpha'})^{1/2}\sqrt{\alpha'}$ for a typical massive string state, and of order $\alpha' m$ for special elongated states in the string spectrum.   For a string of proper size $L$, its worldsheet intersects the horizon along a locus stretched out in $x^+$ by an amount $\Delta x^+=L e^\eta$, where $\eta=\log(dx^+/dx^-)/2$ is its rapidity.  We therefore impose that the separation in $x^+$ as Strings 1 and 2 cross
the horizon is much bigger than the fiducial size of String 1 along the horizon:
\be\label{separate}
 L_1 e^{\eta_1} 
 %+ L_2 e^{\eta_2}
  \ll x_{21,\text{h}}^+.
\ee
 See Figure \ref{figbhws}\ for illustration.  

Before proceeding, it is important to understand the role string theory plays in the effect we are discussing, so let us explicitly contrast it with the behavior of ordinary particles in a black hole background. 
A particle with Schwarzschild energy $E_1$ dropped into the black hole increases the black hole mass $M_{\text{BH}}\sim r_\text{s}/G_\text{N}\sim r_\text{s}/(g_{\text{eff}}^2\alpha')$ by $\Delta M_{\text{BH}}=E_1$,  where $g_{\text{eff}}^2 = g_\text{c}^2/\text{Vol}(M_{\text{internal}})$ is the effective four-dimensional closed string coupling squared.  
 So a later probe feels a Schwarzschild geometry with mass
 \be\label{massdelt}
 M_{\text{BH}}\left(1+O(g_{\text{eff}}^2 \alpha'E_1/r_\text{s})\right)
 \ee    
Thus these corrections are controlled by a factor which for $E_1\ll r_\text{s}/\alpha'$ is parametrically smaller than the closed string coupling (which itself is parametrically smaller than the open string coupling in theories with open strings).

\subsection{Symmetric trajectories}
First let us review the form of the trajectories of massive geodesics in a Schwarzschild black hole. The conserved energy of a radial geodesic satisfies 
\begin{align}
\left(\frac{E}{m}\right)^2=\left(1-\frac{r_{\text{s}}}{r}\right)^2\left(\frac{dt}{d\tau}\right)^2=\left(\frac{dr}{d\tau}\right)^{2}+1-\frac{r_{\text{s}}}{r},
\end{align}
where $m$ is the mass of the particle and $\tau$ is its proper time. For geodesics that start at rest at some $R\gg r_{\text{s}}$, it follows that the energy is
\begin{align}\label{EmR}
E=m\sqrt{1-\frac{r_{\text{s}}}{R}}.
\end{align}
These trajectories have energy $E<m$, as a result of the redshift $\sqrt{-g_{00}}=\sqrt{1-r_{\text{s}}/r}$. The other family of geodesics satisfy $E>m$, and correspond to particles that are sent into the black hole at a nonzero velocity. Using (\ref{boost}), if two particles are sent in on the same trajectory with time separation $\Delta t$, then the ratio of their longitudinal positions at the horizon is 
\begin{align}\label{longratio}
\frac{x^+_{2,\text{h}}}{x^+_{1,\text{h}}}=\exp\left( \frac{\Delta t}{2r_{\text{s}}}\right).
\end{align}
It follows that at large time separations $\Delta t$, the null separation of the particles on the horizon is 
\begin{align}
 x^+_{21,\text{h}}\sim x^+_{1,\text{h}}\exp\left( \frac{\Delta t}{2r_{\text{s}}}\right)\label{nullsep}.
\end{align}
\indent In order to determine the detected longitudinal size of the early string, we will need to compute the momentum of the late infaller and any secondary probes it emits in the near horizon region. Using the conservation laws 
\begin{align}
\frac{E}{m}&=\left(1-\frac{r_{\text{s}}}{r}\right)\frac{dt}{d\tau}\\
1&=-g_{\mu\nu}\frac{\partial x^\mu}{\partial \tau}\frac{\partial x^\nu}{\partial \tau}(r=r_{\text{s}})=2\frac{\partial x^+}{\partial \tau}\frac{\partial x^-}{\partial \tau}(r=r_{\text{s}})\label{momentum},
\end{align}
along with the transformation (\ref{transform}), one finds 
\begin{align}
p^+_{\text{h}}=m\frac{dx^+_{\text{h}}}{d\tau}=\frac{m^2}{4r_{\text{s}}E}x^+_{\text{h}}\label{pminus}.
\end{align}
\indent Let us start with the simplest situation described above where the trajectories of the two strings differ by an overall time translation, i.e. $E_1=E_2=E$ and  $m_1=m_2=m$.  We will also take into account the possibility noted above in (\ref{ellcondgen}) that the detector may be an offshoot of String 2.    
The spreading of String 1 as seen by the detector is then
\be\label{size}
%\Delta X^+_1\sim \alpha' p_{2,\text{h}}^+\sim \frac{\alpha' m^2x_{1,\text{h}}^+}{r_{\text{s}}E}\exp\left( \frac{\Delta t}{2r_{\text{s}}}\right),
\Delta X^+_1\sim \frac{ p_{\text{det},\text{h}}^+}{m_{\text{det}}^2}\sim \left(\frac{ p^+_{\text{det}, \text{h}}}{p^+_{2, \text{h}}}\right)\frac{ m^2 x_{1,\text{h}}^+}{ m_{\text{det}}^2 r_{\text{s}}E}\exp\left( \frac{\Delta t}{2r_{\text{s}}}\right),
\ee
where we used (\ref{longratio}) along with the assumption of large relative boost $\Delta t\gg 2r_{\text{s}}$.  
Here the detector (labeled ``det") may a priori be either String 2 itself, or a secondary probe which we will call String 3.  The latter possibility substantially relaxes the conditions for drama, since one can for example have $m_3\ll m_2$ and $p^+_3$ close to $p^+_2$, effectively removing the $1/(\alpha' m_{\text{det}}^2)$ suppression in the detectable spreading in (\ref{conservative}) relative to the nominal light cone time resolution \cite{lennyspreading}.   
Below in Section \ref{secondary}\ we will carefully analyze the kinematics of such secondary probes to ensure their consistency and lay out the general conditions for a breakdown of EFT.  In the rest of this section we will first apply the estimate (\ref{size}) for $m^2_{\text{det}}= m^2_3\gtrsim 1/\alpha'$ and $p^+_3$ of order $p^+_2$.  (This is a regime for which the saddle point analysis (\ref{Tsaddle}) is marginal, but in the next section, we will see that the basic results are robust when we allow for a larger $m_3$, consistent with the saddle point.)  
In this case,
(\ref{size}) reduces to
\be\label{sizenaive}
\Delta X^+_1\sim \alpha' p_{2,\text{h}}^+\sim \frac{\alpha' m^2x_{1,\text{h}}^+}{r_{\text{s}}E}\exp\left( \frac{\Delta t}{2r_{\text{s}}}\right),  ~~~~~~  m^2_{\text{det}}= m^2_3\gtrsim 1/\alpha', ~~~ p^+_3 \sim p^+_2 .
\ee
This is equivalent to the original estimates in \cite{lennyspreading}, but here for the secondary detector.
We will find an interesting breakdown of EFT very simply from this.     

We will then address the case where $m_{\text{det}}=m_2$ which will not lead to a breakdown of EFT according to the conservative spreading condition (\ref{conservative}).  In the following sections we will generalize our analysis of EFT violations detected by secondary probes, and also address the case of asymmetric trajectories (which will relax the constraints on the mass of String 1).        

Comparing (\ref{nullsep}) with (\ref{sizenaive}),  the condition for the late offshoot string (3) to detect the early string becomes 
simply \be\label{symmcond}
m^2> \frac{r_{\text{s}}E}{\alpha' }.
\ee
Taking $E<m$ and using (\ref{EmR}), this translates into a constraint on the radius $R$ from which the strings 1 and 2 are dropped into the black hole:
\be\label{symmcondR}
m>\frac{r_{\text{s}}}{\alpha'}\sqrt{1-\frac{r_{\text{s}}}{R}}.
\ee
\indent We also need to satisfy the constraint (\ref{separate}). The rapidity of a trajectory with Schwarzschild energy $E$ is 
\begin{align}\label{rapiditydef}
e^{\eta}=\sqrt{\frac{p_{\text{h}}^+}{p_{\text{h}}^-}}=\frac{\sqrt{2}p^+_{\text{h}}}{m}=\frac{\sqrt{2}m}{4r_{\text{s}}E}x^+_{\text{h}},
\end{align}\\
so (\ref{separate}) becomes
\begin{align}\label{fakedrama}
\frac{mL_1}{r_{\text{s}}E}\ll \exp\left(\frac{\Delta t}{2r_{\text{s}}}\right)
\end{align}
\indent One simple way to satisfy the inequality (\ref{symmcondR}) is to take $m>r_{\text{s}}/\alpha'$ and $R\to \infty$, so that the strings are dropped in from far outside the black hole. This is a somewhat strong condition on the mass $m$.  However, the strings can still be taken to be small perturbations to the black hole, since
\begin{align}
\frac{r_{\text{s}}/\alpha'}{M_{\text{BH}}}=g_{\text{eff}}^2\ll 1.
\end{align}
Also, the size of a typical string of this mass is much smaller than the Schwarzschild radius of the black hole background, $ m_2^{1/2}{\alpha'}^{3/4}\sim r_{\text{s}}^{1/2}\alpha'^{1/4}\ll r_{\text{s}}$. 

We should also check that the process is under perturbative control, and dominated by the tree level interaction.  Of course, it is possible that the RMS spreading (\ref{transverse}) and (\ref{deltax+}) is still accurate once loops are taken into account, but the analysis of  \cite{bpst}\cite{usSmatrix} is restricted to tree-level physics.  
To begin with, we should impose that the self-interactions as well as mutual interactions of the strings are controlled by a perturbative expansion, so that loop effects are suppressed relative to the tree level processes.  In order to estimate this, one must take into account the low density of string in a typical massive state:  a super-Planckian mass is perfectly controllable if distributed sufficiently dilutely, as in real-world astrophysics as well as cosmic string theory. For a typical single-string state with mass $m\sim r/\alpha'$ in four dimensions, this implies that \cite{garyjoe}
\begin{align}\label{intcondition}
g_{\text{eff}}\ll \left(\frac{\alpha'}{r_{\text{s}}^2}\right)^{3/4}.
\end{align}
For a pair of strings at rest, the condition would be similar since it involves the local density of interacting string.

One could impose a stronger condition, as follows.  
The center of mass energy squared in the near horizon region is
\begin{align}
s_{\text{h}}\sim 2p_{2,\text{h}}^+p_{1,\text{h}}^-\sim m^2\exp\left(\frac{\Delta t}{2r_{\text{s}}}\right).
\end{align}
%In order to justify the tree-level approximation, this should be 
If we insist that this be smaller than $M_{\text{P}}^2$, for $m\sim r_{\text{s}}/\alpha'$ it would imply that
\begin{align}
g_{\text{eff}}^2\ll \frac{\alpha'}{r_{\text{s}}^2},\hspace{10 mm}\Delta t\ll 2r_{\text{s}}\log\left(\alpha'/(g_{\text{eff}}^2r_s^2)\right ).
\end{align} 
Since this is a stronger condition, it is overly conservative in some regimes since it conflicts with (\ref{intcondition}) \cite{garyjoe}.  However, it can be imposed as a sufficient condition for perturbative control, adjusting $g_{\text{eff}}$ accordingly, and as we will see below this still leads to a breakdown of effective field theory.

\indent We still need to check the condition (\ref{fakedrama}). For a mass of order $m\sim r_{\text{s}}/\alpha'$, the typical state has length $r_{\text{s}}^{1/2}\alpha'^{1/4}$, so (\ref{fakedrama}) becomes
\begin{align}
\left(\frac{\alpha' }{r_{\text{s}}^2}\right)^{1/4}\ll \exp\left(\frac{\Delta t}{2r_{\text{s}}}\right).
\end{align}
This is automatically satisfied for macroscopic black holes. \\
\indent Another strategy for satisfying (\ref{symmcondR}) is to take $R\to r_{\text{s}}$, which leads to the inequality
\begin{align}
R-r_{\text{s}}<\frac{(\alpha' m)^2}{r_{\text{s}}}.
\end{align}
In order to exhibit a nontrivial breakdown of effective field theory, the strings must be dropped in from $R-r_{\text{s}}>\sqrt{\alpha'}$, which implies the condition
\begin{align}
m^2>\frac{r_{\text{s}}}{\alpha'^{3/2}}.
\end{align}
This is a weaker condition than $m>r_{\text{s}}/\alpha'$.\\
\indent Next let us discuss how the results are modified if we do not make use of the secondary probe, String 3.
 %instead take the conservative spreading formula (\ref{conservative}). 
If $m_{\text{det}}=m_2 = m$ in (\ref{size}),  the condition for drama for $m^2>1/\alpha'$ becomes $r_{\text{s}}E<1$. For a trajectory with $E<m$, this implies that
\begin{align}
m\sqrt{1-r_{\text{s}}/R}<\frac{1}{r_{\text{s}}}\hspace{5 mm}\Rightarrow \hspace{5 mm}R-r_{\text{s}}<\frac{R}{m^2r_{\text{s}}^2}<\frac{\alpha' R}{r_{\text{s}}^2}\ll \sqrt{\alpha'},
\end{align}
where we used that $r_{\text{s}}^2\gg \alpha'$. As a result, this trajectory requires acceleration at greater than the string scale, and hence fails to exhibit a breakdown of effective field theory. In the processes involving secondary probes, there is drama as we have seen.  We will next spell out more completely the consistency conditions for the processes involving secondary probes.  

%there will be violations of effective field theory even if we do apply the conservative condition (\ref{conservative}).
% order for the detectable spreading to be limited to the range $p_{2-}/m^2$ rather than $p_{2-}\alpha'$, as in \S\ref{puzzles}.  

\subsection{Drama from secondary probes:  more details on the kinematics}\label{secondary}

In applications to black hole (thought) experiments, the late infaller may perform detailed experiments to access the physics in the near horizon region. For example, 2 can send a pulse 
%--one that may even reach $r=\infty$ and 
that detects the longitudinal spreading of String 1. We can model this as a decay process, $2\to 2'+3$.   This may also include the possibility an outgoing probe that could detect the spreading, in addition to probes which fall into the black hole. \\
\indent Let us warm up with the simplest case, where 3 is an outgoing massless string with Rindler energy $p_3^+$. %We take $m_{2}-m_{2'}\ll m_2$ so that the detector does not shoot off a large portion of its mass during the experiment. 
Kinematically this process requires the energy-momentum conservation conditions (with no transverse momentum for simplicity)\footnote{Again we note, however, that we do not have a precise calculation along the lines of \cite{bpst}\ for this case, since there the derivation of the saddle point in $T$ described above in (\ref{Tsaddle}) depended on the nonzero transverse momentum transfer $p_\perp^2\gg 1/\alpha'$.  We will see shortly that the results are similar for a significant range of $m_3^2$ (equivalently $p_{\perp, 3}^2$).}  
\begin{align}
p_{2,\text{h}}^+=p^+_{2',\text{h}}+p_{3,\text{h}}^+,\hspace{15 mm}\frac{p_{2,\text{h}}^+}{m_2^2}=\frac{p_{2',\text{h}}^+}{m_{2'}^2}.
\end{align}
Solving for $p_{3,\text{h}}^+$, one finds 
\begin{align}
p_{3,\text{h}}^+=p_{2,\text{h}}^+\left(1-\frac{m_{2'}^2}{m_2^2}\right).
\end{align}
The condition for the probe to detect String 1 is then 
\begin{align}
x_{21,\text{h}}^{+}\sim \frac{r_{\text{s}}E_2p_{2,\text{h}}^+}{m_2^2}< \alpha'p_{3,\text{h}}^+.
\end{align}
%$m_{2}-m_{2'}\ll m_2$, as well as 
%$E_2\approx m_2$ to avoid strong acceleration, 
This condition is satisfied when
\begin{align}\label{lightraymasses}
m_2\left(1-\frac{m_{2'}^2}{m_2}\right)>\frac{r_{\text{s}}}{\alpha'}\frac{E_2}{m_2}.
%,\hspace{10 mm}m_{2}-m_{2'}>\frac{r_{\text{s}}}{\alpha'}.
\end{align}
Note that this result is still true even given the conservative estimate discussed above around equation (\ref{conservative}), since String 3 is massless and hence traverses a spreading radius within the light cone resolution time.  For the case $m_{2'}$ of order $m_2$ (but not tuned to agree precisely), this reproduces (\ref{symmcond}) as anticipated above.
It applies equally well to a slightly massive secondary probe with $m_3 \sim 1/\sqrt{\alpha'} \ll p_3^+$.  \\
\indent More generally, the secondary probe 3 could be a massive string state, with mass $m_3^2>1/\alpha'$. The conservative condition for drama is then
\be\label{probecond}
x^+_{21,\text{h}}\sim \frac{r_\text{s}E_2p^+_{2,\text{h}}}{m_2^2}< \frac{p^+_{3,\text{h}}}{m_3^2} 
\ee
In order for the probe to detect the spreading that 2 could not (assuming the conservative estimate (\ref{conservative})), we must have
\be\label{epsell} 
\frac{p^+_{3,\text{h}}}{m_3^2} \equiv \frac{1}{\epsilon} \frac{p^+_{2,\text{h}}}{m_2^2}, ~~~~~~~ \epsilon \ll 1. \ee
We will now determine the conditions on our trajectory parameters in order to obtain sufficiently small $\epsilon$ for this detection, taking $R\gg r_{\text{s}}$ as above to avoid strong acceleration. Conservation of momentum implies that
\begin{align}\label{laserconservation}
p^+_{2,\text{h}} &= p^+_{2',\text{h}}+p^+_{3,\text{h}} \\
\frac{m_2^2}{p^+_{2,\text{h}}}&=\frac{m_{2'}^2}{p^+_{2',\text{h}}}+\frac{m_3^2}{p^+_{3,\text{h}}}\label{laserconservation2}.
\end{align}   

Plugging (\ref{epsell}) into the conditions (\ref{laserconservation})-(\ref{laserconservation2}) and solving for $m_{2'}^2$, we obtain
\be\label{solpl}
\frac{m_{2'}^2}{m_2^2}=(1-\epsilon)\left(1-\frac{1}{\epsilon}\frac{m_3^2}{m_2^2}\right).
\ee
In order for this to give a consistent solution for $m_{2'}$, we require the right hand side to be positive, so in our regime $m_3^2>1/\alpha'$ we have
\be\label{masseps} 
\frac{1}{\alpha'}< m_3^2 < \epsilon m_2^2 ~~~\Rightarrow ~~~ m_2^2>\frac{1}{\alpha'\epsilon}.
\ee
%where the first inequality holds in the regime we specified above (\ref{probecond}).  
Finally, we can determine the required $\epsilon$.  The drama condition (\ref{probecond}) requires $r_{\text{s}}E_2< 1/\epsilon$.  Putting this together with (\ref{masseps}) we have
\be\label{nextineq}
\sqrt{\frac{r_{\text{s}}^2}{\alpha'\epsilon}}< m_2 r_{\text{s}} < \frac{1}{\epsilon}\frac{m_2}{E_2}
\ee
This in turn implies
\be\label{finalineq}
\epsilon <\frac{\alpha'}{r_{\text{s}}^2}\left(\frac{m_2}{E_2}\right)^2 ~~~ \Rightarrow ~~~ m_2 > \frac{r_{\text{s}}}{\alpha'}\frac{E_2}{m_2}.
\ee
%For $E_2\approx m_2$, 
This is the same condition as above, (\ref{lightraymasses}).

Let us make two more basic consistency checks.  The center of mass energy squared is
\be\label{ssecondary}
s_{\text{h}}\sim 2 p^+_{2,\text{h}}p_{1,\text{h}}^- \sim m_2^2 \exp\left(\frac{\Delta t}{2r_{\text{s}}}\right).
\ee
This can easily be smaller than $M_\text{P}$, a sufficient condition to avoid backreaction and maintain perturbativity, as noted above.
With the secondary probe $3$ making the detection, we should also check that $n_{\text{max}}$ is still large; this is
\be\label{nmaxell}
n_{\text{max}, 3} = -2 \alpha' p_{3,\text{h}}\cdot p_{1,\text{h}}\sim \left(\frac{1}{\epsilon} \frac{m_3^2}{m_2^2}\right)\alpha'p^+_{2,\text{h}}p_{1,\text{h}}^-\sim  \left(\frac{1}{\epsilon} \frac{m_3^2}{m_2^2}\right) \alpha' m_2^2\exp\left(\frac{\Delta t}{2r_{\text{s}}}\right).
\ee
The factor in parentheses must be less than one by (\ref{masseps}), but this still leaves a wide window with $n_{\text{max}}\gg 1$. 
%The case where 3 is taken to be massless works similarly.

It would be interesting to generalize this analysis to the potential detection of String 1 by emitted Hawking radiation.  We will leave a careful analysis of that to future work.  

\subsection{Black hole as accelerator and the breakdown of EFT (it's not just Rindler dynamics)}\label{BHaccelerator}

The expression (\ref{ssecondary}) illustrates one of the main features of the black hole problem  that we reviewed in the introduction:   although the Schwarzschild geometry is weakly curved, it generates an enormous center of mass energy in the near horizon region, motivating a more careful study of the breakdown of EFT in string theory.  The evolution of the trajectories with energy $E=m$, separated by a Schwarzschild time $\Delta t$, accelerates a system of a given external center of mass energy squared
\be\label{sextsec}
s_{\text{ext}} \sim m^2 
\ee
far outside the black hole to one with the large center of mass energy squared (\ref{ssecondary}), boosted exponentially in the Schwarzschild time.  Note that this effect is different from the large accelerations and direct collisions generated in \cite{Kerraccel}, an effect that may be interesting to combine with our study here.      

This feature is absent in flat spacetime physics; the drama we obtain in this way is very naturally generated by the black hole.  Of course there are manifestations of string spreading effects in the flat space S-matrix, as is well known in the case of transverse spreading and which we analyze for evidence of longitudinal spreading in \cite{usSmatrix}.  In that context, the large center of mass energy is introduced from the start.  In the application to black hole physics,  the black hole functions as a very effective accelerator, generating the large relative boost required for string-theoretic effects to introduce a breakdown of effective field theory.  This occurs despite the weak curvature of the Schwarzschild geometry in the exterior and near-horizon regions.  In this sense, the black hole naturally produces conditions leading to the breakdown of effective field theory, given that it is UV completed by string theory.        

\subsection{Causality}

It is also important to emphasize that the effect we are finding is perfectly consistent with causality; it is just nonlocal.  According to \cite{lennyspreading}\ as reviewed above, a string is generally spread out.  This is only detected when a second probe has sufficient resolution to interact with the high ($\sim n_{\text{max}}$) modes of the string.   The late detector does not cause the spreading of the early string (or vice versa), since the strings are always spread out.  

As mentioned above, there is a caveat to this last statement:  we are treating the string that arrives in the near horizon region is in a standard flat space single-string state, as entered into the calculations \cite{lennyspreading}\ and above.  There is a logical possibility that the curvature of the geometry obstructs this during the infall process.

\subsection{Asymmetric trajectories}\label{asymmetric}

Let us now allow for distinct trajectories, $E_1\ne E_2, m_1\ne m_2$.  
  
For trajectories with $E_j<m_j$, with string $j$ dropped in from rest at $r=R_j$, we must require 
\be\label{accelcond}
R_j-r_{\text{s}}\gg \sqrt{\alpha'} ~~ \Rightarrow ~~ \left(\frac{m_j}{E_j}\right)^2-1\ll \frac{r_{\text{s}}}{\sqrt{\alpha'}} ~~~~~ (E_j<m_j)
% ~~ \Rightarrow E_2\gg \frac{1}{r_{\text{s}}} m_2\sqrt{\alpha'}
\ee
in order to avoid string-scale acceleration.  Here we used the relation (\ref{EmR}). 

As with the symmetric trajectories, let us start with the simplest case analogous to (\ref{sizenaive}), with
$m_{\text{det}}= m_3\sim 1/\alpha',  p^+_3 \sim p^+_2$.
We have the condition for String 2's offshoot, String 3, to detect String 1 via longitudinal spreading:
\be\label{asymmcond}
x_{21,\text{h}}^+ < 
%c_L
\frac{\alpha' m_2^2}{E_2 r_{\text{s}}}x_{2,\text{h}}^+  ~~~~~~~~ m_{\text{de}t}=m_3\sim 1/\alpha', ~~~ p^+_3 \sim p^+_2
\ee
%Here again $c_L$ is a constant of order 1, with the amplitude of the Gaussian-distributed spreading expected to degrade like $e^{-c_L^2}$.
 %Let's write (\ref{asymmcond})
In general, it is useful to work in terms of the near-horizon rapidities (\ref{rapiditydef}), which are determined by
%Using (10) from ESnotes we get
\be\label{etaexplicit}
\eta=\frac{1}{2}\left(\frac{R}{r_{\text{s}}}-1 +\sqrt{R/r_{\text{s}}-1}(2+R/r_{\text{s}})\tan^{-1}\sqrt{R/r_{\text{s}}-1} \right)+t_R/2r_{\text{s}},
\ee  
for the $E<m$ trajectories in which a string is dropped from rest at $r=R$ at time $t=t_R$.  We note that as $R\to\infty$, at fixed $t_R$ the rapidity $\eta$ at the horizon blows up.  This is because it takes a longer time to fall in from further out.  But correspondingly if we take $t_R$ sufficiently negative then $\eta$ can remain finite at the horizon.  In terms of this, our condition for drama (\ref{asymmcond}) combined with the inequality (\ref{separate}) takes the form
\begin{align}\label{asymmcondetalowm}
 L_1 e^{\eta_1}
 %+L_2 e^{\eta_2} 
\ll  r_{\text{s}} \left(\frac{E_2}{m_2}e^{\eta_2}-\frac{E_1}{m_1}e^{\eta_1}\right) < \alpha' m_2 e^{\eta_2}.
%e^{\eta_1-\eta_2}+\frac{m_1}{E_1(2\sqrt{e}r_{\text{s}})}L_1 e^{\eta_1-\eta_2} ~ < ~ \frac{E_2}{m_2}\frac{m_1}{E_1} ~  < ~ e^{\eta_1-\eta_2} &+&\left( \frac{ E_2m_1}{m_2 E_1}\right)\frac{m_2^2\alpha'}{2\sqrt{e}\  r_{\text{s}} E_2} ~~~~~~~~~(m_2^2\alpha'<1).
\end{align}

We obtain two more conditions as follows.
 % In order for the resolution of the detector to be fine enough for the results of Section 2 to be applicable, we need the center of mass energy of the two strings in the near horizon region to exceed the string scale. 
Using $p^+=m\sqrt{dx^+/dx^-}/\sqrt{2}=m e^\eta/\sqrt{2}$ and $p^-=m^2/(2p^+)$, we can write the squared center of mass energy in the near-horizon region as
\begin{align}\label{snearhorizon}
s_{\text{h}}\sim -2p_{1,\text{h}}\cdot p_{2,\text{h}}\sim m_1 m_2\left(e^{\eta_2-\eta_1}+e^{\eta_1-\eta_2}\right) \approx  m_1 m_2 e^{\eta_2-\eta_1} .
\end{align}
%For $m$ satisfying (\ref{bigmass}), this is automatically bigger than the string scale. 
The condition $n_{\text{max}}\gg 1$  (\ref{nmaxlarge}) requires $s\gg 1/\alpha'$. 
 But another lower bound arises from the following considerations.   Let us compare this near horizon $s_{\text{h}}$ to that we start with externally, outside the black hole.  
The latter is 
\begin{align}\label{sexternal}
s_{\text{ext}}\sim -2p_{1\text{ext}}\cdot p_{2\text{ext}}\sim m_1 m_2\left(e^{\eta_{2,\text{ext}}-\eta_{1,\text{ext}}}+e^{\eta_{1,\text{ext}}-\eta_{2,\text{ext}}}\right) = m_1m_2\left(\frac{E_1 m_2}{m_1 E_2}+\frac{E_2 m_1}{m_2E_1}\right).
\end{align}
If we wish to ensure that the drama we obtain is particular to the black hole problem, as opposed to just exhibiting string-theoretic effects which would also be present in Minkowski spacetime, we should insist that
\be\label{BHdrama}
s_{\text{ext}}\ll s_{\text{h}} ~~~~~ {\rm (for~ black~hole-induced~ drama)}
\ee
Of course in flat spacetime, string theory goes beyond EFT, as can be seen in scattering amplitudes (including those exhibiting longitudinal nonlocality plausibly related to longitudinal string spreading \cite{usSmatrix}).  But for our present purposes, we would like to focus on the breakdown of EFT arising in black hole physics as a result of the evolution of the trajectories in the black hole geometry, as discussed above for symmetric trajectories in Section \ref{BHaccelerator}.    

This gives a lower bound on $s$.  On the other hand, we can impose that the center of mass energy not exceed the Planck scale, a sufficient condition for perturbative control (although as discussed above this is sometimes overly conservative, as sufficiently dilute super-massive objects can be described reliably in perturbation theory without strong gravity or string loops). These two conditions require the window (taking $\eta_2>\eta_1$)
\begin{align}\label{Planckwindow}
\frac{E_1 m_2}{m_1 E_2}+\frac{E_2 m_1}{m_2E_1}\ll e^{\eta_2-\eta_1} \ll \frac{M_\text{P}^2}{m_1m_2}
%\Delta t<r_{\text{s}}\log \left(\frac{M_{\text{pl}}}{m}\right),
\end{align}
It is useful to define the quantity
\be\label{tHdef}
\Delta t_{\text{h}} \equiv 2 r_{\text{s}} \log(x^+_{2,h}/x^+_{1,h}) =2 r_{\text{s}}\left(\eta_2-\eta_1+\log\left(\frac{E_2 m_1}{m_2 E_1}\right)\right)
\ee
which is the difference in Schwarzschild times between the horizon crossing of the early and later infallers. 
For the symmetric trajectories, this is proportional to the relative boost, while in the asymmetric case it is offset from this by the logarithmic term.  This difference $\Delta t_\text{h}$ is finite (generically) in our system, although of course the Schwarzschild time coordinate $t$ diverges near the horizon.  As reviewed in \cite{JoeBHcomp}, nice slice coordinates can be formuated which are smooth across the horizon and asymptote to $t$ at infinity, with (\ref{tHdef}) the time difference in these coordinates.  
The condition (\ref{BHdrama}) (the first inequality in (\ref{Planckwindow})) implies
\be\label{tHlarge}
\Delta t_\text{h} \gg 2 r_{\text{s}}.  ~~~~~ {\rm (for~ black~hole-induced~ drama)}
\ee
That is, when the drama is generated by the black hole, it sets in beyond the timescale $\sim 2 r_{\text{s}}$ regardless of whether the trajectories are symmetric or asymmetric.   We will discuss this further below in Section \ref{firewall}.\\
\indent Given the condition (\ref{tHlarge}), the condition for drama (\ref{asymmcond}) is satisfied when
\begin{align}\label{m2asymm}
m_2^2>\frac{r_{\text{s}}E_2}{\alpha'}.
\end{align} 
This is identical to (\ref{symmcond}), but without a constraint on $m_1$. This strengthens the significance of amplitude of the string-theoretic effect relative to the pure general-relativistic effects in (\ref{massdelt}), which become more important as the mass of the early perturbation increases. One could also consider the case where (\ref{tHlarge}) is not satisfied, so that (\ref{asymmcond}) becomes
\begin{align}
m_2^2>\frac{E_2r_{\text{s}}}{\alpha'}\left(1-\exp\left(-\frac{\Delta t_{\text{h}}}{2r_{\text{s}}}\right)\right).
\end{align}
This relaxes the bound on $m_2$ by the factor in parentheses. However, the second term is negligible if we require that the black hole generates the large relative boost.\\
\indent As in the symmetric case, the asymmetric trajectories do not give drama for the conservative spreading estimate (\ref{conservative}) without a secondary probe. It is straightforward to generalize to asymmetric trajectories the detailed analysis of the kinematics with secondary probes, verifying the condition (\ref{m2asymm}). 

\section{Strings and the firewall paradox:  the UV sensitivity of horizon dynamics}\label{firewall}

The work \cite{firewalls}\ gives a precise argument confirming the longstanding problem that effective field theory near the horizon of a late black hole is incompatible with unitary quantum mechanics, despite previous arguments to the contrary, and speculates that the conflict is resolved via a sharp `firewall' developing just inside the horizon.\footnote{For a sample of additional references also sharpening the black hole information problem in important ways, see \cite{preAMPS}\ and of course \cite{hawking}.}   
Our analysis combined with \cite{usSmatrix}\ has provided strong evidence for the breakdown of effective field theory for probes falling across the horizon much later than early matter, if the latter includes strings.   
Given this, among the incompatible postulates delineated by AMPS, string theory appears to automatically violate the validity of EFT outside the horizon, as well as the `no drama' condition for an infaller.   

This provides a good candidate for the dynamics behind the required breakdown of effective field theory anticipated in \cite{firewalls}.  However it is worth emphasizing that the effect we found so above is not precisely a `firewall', as the effect we find in string theory sets in as soon as the strings reach the near-horizon region, once the large relative boost is generated. (However, as we will note momentarily, there is potential for additional effects if the Hawking radiation can detect the spreading.)    It is also not obtained from any straightforward generalization of EFT to a non-local version \cite{NVNL}, in that the effect is clearly dependent on the basic microphysics in string theory -- the extended nature of the fundamental strings.   Our effect shares features of both of these scenarios, but arises in a somewhat clearer framework which produces a more specific mechanism for the breakdown of EFT.  

It is worth emphasizing that we established a breakdown of EFT despite at various points imposing potentially overly conservative conditions, and without invoking additional effects such as the sum over early matter strings and the Hagedorn density of available final string states.  Kerr geometries also introduce additional acceleration effects, which we have not yet applied to string spreading.   Finally, we stayed away from the regime where loops might become important, and we do not know whether the effect persists in some form in the presence of stronger interactions, and if so whether it enhances the effect or suppresses it. 

We believe that this breakdown of EFT strongly motivates further analysis of black hole thought experiments with string-theoretic effects included, in order to determine whether residual paradoxes remain, as well as to explore information transfer in string theory.  For now, we will make some preliminary remarks.  

First, it is worth noting that one may be able apply our estimate for detectable spreading by the light secondary probe, String 3 above, to outgoing Hawking particles in the vicinity.  A symmetric arrangement of early and late Hawking particles would be analogous to our two-body symmetric trajectories  above, giving no detection.  But the interaction between early matter and late outgoing Hawking particles is more analogous to the setup with the secondary probe, and could be sensitive to spreading for some range of parameters.  It is also interesting to ask whether the early string could disrupt the vacuum entanglement between modes, if one member of the entangled pair can detect the spreading.  If such an effect occurs, it could produce a firewall in the spirit of \cite{firewalls}.  However, vacuum fluctuations are not long-lived and do not immediately fit into our analysis.  Presumably they should not be disrupted in pure flat spacetime physics, which should constrain the effect in the near horizon region of the black hole. However, the mining processes involving high partial waves described in \cite{firewalls}\ might introduce real high-energy quanta that can detect the spreading of the early matter. We will leave this very interesting analysis of the effect of longitudinal string spreading on the Hawking emission to future work. 

In any case, let us consider the timescale at which black hole-induced drama sets in.  We have considered a process where String 1 falls into a pre-existing black hole horizon, and found that after a time $\Delta t_\text{h} \sim 2 r_{\text{s}}$ (\ref{tHlarge}), an infalling String 2 -- more precisely its offshoot String 3 -- can detect the longitudinal spreading of String 1.  
%In order to compare this to the   
%other timescales considered in black hole formation and evaporation, we should focus on the contribution to them arising from the incremental contribution of string 1 to the black hole entropy.   
Our string 1 is ultimately intended as a proxy for early matter forming the black hole, but our analysis has focused on the toy problem where it falls into a large Schwarzschild black hole.

One timescale of interest is the scrambling time \cite{scrambling}, which is much shorter than the Page time \cite{Pagetime}.  For the original black hole plus our perturbation by String 1, this time is $t_{\text{scrambling}}\sim r_{\text{s}} \log(r_{\text{s}}^2 M_\text{P}^2)\sim \beta\log S$ where $\beta$ is the inverse temperature of the black hole and $S$ its entropy.  From (\ref{Planckwindow}), the condition for control of perturbative string theory, combined with (\ref{tHlarge}), the condition that the black hole induce the drama, we have
\bea\label{deltatscrambling}
2 r_{\text{s}}\ll \Delta t_\text{h} \ll 2 r_{\text{s}}\left(\log(r_{\text{s}}^2 M_\text{P}^2) -
%\log(r_{\text{s}}^2 m_1 m_2)\right) 
 \log(r_{\text{s}}^2 m_2^2 E_1/E_2)\right) 
%t_{scrambling}/\beta &\equiv& t_{scr, 0}/\beta + \Delta ( t_{scr} \nonumber \\
%&\simeq& r_{\text{s}}\log(M_P r_{\text{s}}) + \Delta r_{\text{s}}(\log(r_{\text{s}} M_P)+1) \\
\eea
For symmetric trajectories, the upper bound on this timescale is always less than the scrambling time, but for extremely asymmetric ones there is a wider range of possibilities that satisfy the condition for perturbative control.  
The thought experiments require drama by the Page time $\sim r_{\text{s}}^3M_\text{P}^2$, and we find it well below that timescale.  

Let us make one final remark about this window.  The upper bound in (\ref{deltatscrambling}) ensures perturbative string control, and it is interesting to consider what happens beyond that timescale.  The fact that in its regime of control perturbative string theory automatically provides drama suggests that this will continue to happen in the non-perturbative regime, for any consistent UV-completion of gravity.  That is, it is natural to expect that any consistent theory of quantum gravity will produce the required dynamics, much as it should produce an accurate microstate count for black holes.  Our analysis of this question within perturbative string theory gives a controlled example of what is likely a more general phenomenon.  

In \cite{lennyspreading}, the combination of transverse and longitudinal spreading provided an appealing picture of information spreading onto a `stretched horizon'.  In that work, aimed at providing evidence for the idea of black hole complementarity, it was suggested that a late infaller would not detect anything unusual while crossing the horizon.   Although we find instead that the same underlying physics -- the longitudinal spreading -- leads to drama for the late observer, it may be that the information transfer in the system proceeds along the lines outlined in \cite{lennyspreading}.  We will leave it for future work to flesh this out.  

Another  application of spreading is to thermal AdS/CFT dynamics, as 
analyzed recently in \cite{ShenkerStanford}, where the transverse spreading played an important role.  Longitudinal spreading, if present, was argued to be consistent with what is known about thermalization.  However, it would be interesting to pursue further the dynamics of thermalization contained in the detection of 1 by 2 and its secondary probes in our system.  

\subsection{Observational tests??}

Moving beyond the realm of thought experiments, it is interesting to consider the possibility that this effect could yield signatures in astrophysical observations, either in fundamental string theory or hadron physics.\footnote{We note that although question-titles are always answered in the negative, this one is a double negative.}
This was explored recently in \cite{SteveObservations}\ in the context of an approach to modeling potential nonlocalities in a generalized effective field theory description \cite{NVNL}.   This work proposed to test the resulting deviations from general relativity using the Event Horizon Telescope \cite{EHT}; see \cite{Solvay}\ for a recent overview of observational black hole physics.  

Since in the perturbative string regime, we find a short timescale for drama, it may be worthwhile to examine potential observational signatures of our mechanism (after generalizing to the Kerr geometry).  
Of course, the main challenge in this program involves disambiguating apparent deviations from ordinary astrophysical nonlinearities, and it is an interesting program of research in itself to develop methods for controlling this and providing realistic forecasts for  the resulting sensitivity.    Since most previous approaches to exploiting this upcoming data refer to other types of theoretical corrections to general relativity, it may be interesting to develop more specific strategies aimed at testing the effects of near-horizon drama.  This is a challenging problem in itself; we will leave a joint analysis of the combined constraints from real and thought experiments for future work.

\section{Consistency with constraints from cosmology}

Strong relative boosts also arise in cosmological horizons between early and late-infalling trajectories.
Having argued that string spreading produces a breakdown of effective field theory leading to horizon drama, we must check its basic consistency with cosmological data.  

The standard model of cosmology describes structure formation via a process of frozen-out super-horizon modes continually entering the observable horizon.  In inflationary theory, these modes originate as quantum fluctuations of the scalar perturbation.   
In the early universe, there is compelling evidence from the CMB for the presence of super-horizon perturbations \cite{SpergelZaldarriaga}, and the standard $\Lambda$CDM model is a good fit to cosmological data
%, with a reduced $\chi^2$ of 1.03-1.04 
\cite{Planck}.  Although the data still admit the possibility of deviations such as tensor modes, residual structure in the power spectrum or higher order correlations, quantitatively these are small perturbations even though they would have important implications.  Any dramatic departure from the standard model would be immediately ruled out.  

In the late universe,  matter is distributed homogeneously along approximately flat spatial slices of the spacetime geometry.  In the rest of this section, we will derive the conditions for cosmology-induced drama in this setting.  As for black holes, de Sitter spacetime can function effectively as an accelerator, generating large near-horizon center of mass energy.  The kinematics in cosmology is somewhat different, involving a homogeneous distribution of matter stretching across each observer horizon, as opposed to the above setup with the early and late infallers into the black hole event horizon.  Nonetheless much of the analysis is similar, as we will see shortly.     

%\subsection{Kinematics}

%* homogeneous distribution of `galaxies' in dS leads to detectable spreading 

\begin{figure}
\begin{center}
\begin{tikzpicture}[scale=2.5]
\draw (-1,-1) -- (-1,1) -- (1,1) -- (1,-1) -- (-1,-1);
\draw[color=PineGreen] (1.1,0) node{ $t$};
\draw[color=PineGreen]  (1.1,-.9) -- (1.1,-.1);
\draw[color=PineGreen ,->] (1.1,.1) -- (1.1,.9);
\draw[color=PineGreen ] (1.15,-1) node{\small{$-\infty$}};
\draw[color=PineGreen ] (1.15,1) node{\small{$\infty$}};
\draw[color=RoyalPurple] (-1.1,0) node{$\gamma$};
\draw[color=RoyalPurple]  (-1.1,-.9) -- (-1.1,-.1);
\draw[color=RoyalPurple ,->] (-1.1,.1) -- (-1.1,.9);
\draw[color=RoyalPurple]  (-1.15,-1) node{\small{$-\frac{\pi}{2}$}};
\draw[color=RoyalPurple]  (-1.15,1) node{\small{$\frac{\pi}{2}$}};
\draw [color=RoyalBlue](0,1.1) node{$\theta$};
\draw[color=RoyalBlue] (.1,1.1) -- (.9,1.1);
\draw[color=RoyalBlue,->] (-.1,1.1) -- (-.9,1.1);
\draw[color=RoyalBlue] (1,1.1) node{$0$};
\draw[color=RoyalBlue] (-1,1.1) node{$\frac{\pi}{2}$};
\draw (0,0) -- (1,1);
\draw (0,0) -- (1,-1);
\draw[color=BurntOrange] (.4,.55) node{$x^+$};
\draw[color=BurntOrange] (.4,-.55) node{$x^-$};
\draw[dashed] (.65,-1) -- (.65,1);
\draw[dashed] (.9,-1) -- (.9,1);
\draw (.58,0) node{\small{$\theta_1$}};
\draw (.83,0) node{\small{$\theta_2$}};
\draw[color=BurntOrange] (.8,-1) -- (.43,-.63);
\draw[->,color=BurntOrange] (.3,-.5) -- (-.05,-.15);
\draw[<-,color=BurntOrange] (.75,.95) -- (.43,.63);
\draw[color=BurntOrange] (.25,.45) -- (-.05,.15);
\end{tikzpicture}
\end{center}
\caption{Trajectories 1 and 2 in the late de Sitter universe.  For small values of the global spatial coordinate $\theta$, the trajectories fall across the indicated observer horizon at a late global time, so that the spatial slices are nearly flat as in our observed universe.  Within that regime, the hierarchy $\frac{\theta_2}{\theta_1}\ll 1$ leads to a large relative boost at the horizon, generated by the cosmological background.  \label{dSfig}  
}
\end{figure}
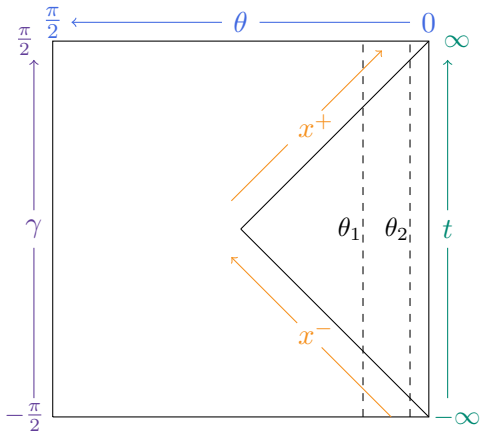

We are interested in a  spatially homogeneous distribution of matter, including our source string 1 and detector 2.  For simplicity, let us consider de Sitter spacetime in global coordinates 
\be\label{dSglobal}
ds^2= \frac{L^2}{\cos^2\gamma}\left(-d \gamma^2 + d\theta^2 +\sin^2\theta\, d\Omega_2^2\right) 
\ee
with $-\pi/2 < \gamma <\pi/2$.  Let us distribute our matter uniformly along the spatial directions here, but consider the system at very late times (so that the positive curvature of these spatial slices is unimportant -- we use these slices just for simplicity).  
%(FIG:  dS penrose diagram, with static patch marked observer at $\theta=0$).

de Sitter spacetime has observer-dependent horizons.  Let us consider an observer at $\theta=0$, and suppose that
the source string 1  at $\theta=\theta_1$ and detector 2 at $\theta=\theta_2$ are falling across that observer's horizon at late times, $\gamma \approx \pi/2$ (see Figure \ref{dSfig}).  This requirement that they cross the horizon at late times implies $\theta_1,\theta_2\ll 1$, and we will specialize to this regime after deriving the fixed-$\theta$ trajectories below.  

It is useful to work with static coordinates adapted to this observer's patch, along with Kruskal-like coordinates for de Sitter spacetime.\footnote{See \cite{StromingerdS}\ for a pedagogical description of coordinate systems in dS.}  These are given by
\be\label{static}
ds^2 = -\left(1-\frac{r^2}{L^2}\right)\, dt^2 + \frac{dr^2}{1-\frac{r^2}{L^2}} + r^2 \, d\Omega_2^2
\ee          
and
\be\label{KruskaldS}
ds^2=\frac{1}{(1-x^+x^-/L^2)^2}\left(-4dx^+dx^-+L^2(1+x^+x^-/L^2)^2\,d\Omega_2^2\right)
\ee     
with the relation
\be\label{xr}
x^\pm =\pm L e^{\pm t/L}\sqrt{\frac{L-r}{L+r}}.
\ee      
It follows that the ratio of the $x^+$ coordinates of 1 and 2 at the horizon $r=L$ grows exponentially with the time separation,
\begin{align}
x^+_{2,\text{h}}=\exp\left(\frac{\Delta t}{L}\right)x^{+}_{1,\text{h}}.
\end{align}
\indent The conserved energy in the static patch is 
\begin{align}
\left(\frac{E}{m}\right)^2=\left(1-\frac{r^2}{L^2}\right)^2\left(\frac{dt}{d\tau}\right)^2=\left(\frac{dr}{d\tau}\right)^2+1-\frac{r^2}{L^2},
\end{align}
so trajectories that start at rest at $r=R$ satisfy
\begin{align}
E=m\sqrt{1-\frac{R^2}{L^2}}.
\end{align}
The analogue of (\ref{pminus}) becomes
\begin{align}
p^+_{\text{h}}=m\frac{dx^+}{d\tau}(r=L)=\frac{m^2}{2EL}x^+_{\text{h}}.
\end{align}
\indent Given these kinematics, we can proceed similarly to our above analysis of the black hole case, taking symmetric trajectories for simplicity. The condition for detectable spreading with a secondary probe is
\bea\label{spreaddS}
x^+_{21,\text{h}} < \alpha' p_{2,\text{h}}^+,
\eea  
and the condition that the cosmological background induce the drama by generating a large relative boost is
\be\label{tHlargedS}
\frac{x^+_{2\text{h}}}{x^+_{1\text{h}}} \sim \exp\left(\frac{\Delta t}{L}\right) \gg 1.
\ee
Putting these together, (\ref{spreaddS}) then becomes
\be\label{mtwoconddS}
m^2>\frac{LE}{\alpha'}.
\ee
For $E\sim m$, this is solved by $m>L/\alpha'$. This is analogous to the condition $m>r_{\text{s}}/\alpha'$ in the black hole case. It is straightforward to repeat this analysis for secondary probes and asymmetric trajectories, and one again finds the condition $m_2>L/\alpha'$. 

\indent For the reasons discussed above, we are interested in strings that are sent in at small $\theta$ in global coordinates. For a trajectory at fixed $\theta$ in (\ref{dSglobal}), one finds
\be\label{tofr}
\sinh (t/L) = \frac{1}{\sin\theta}\sqrt{\frac{r^2-L^2\sin^2\theta}{L^2-r^2}}
\ee
This solution is straightforward to check using the energy conservation condition in the static patch, which is equivalent to
\be\label{rdotE}
\frac{dr}{dt}=\left(1-\frac{r^2}{L^2}\right)\sqrt{1-\frac{m^2}{E^2}\left(1-\frac{r^2}{L^2}\right)}
\ee
Our trajectories trace back to ones that start from rest at $t=0$ from radial position $r=R=L\sin\theta$, with 
\be\label{ERdS}
\frac{E}{m}=\sqrt{1-\frac{R^2}{L^2}}=  \cos\theta.       
\ee
Therefore the trajectories with small $\theta$ correspond to geodesics in the static patch with $E\sim m$, which requires $m_2>L/\alpha'$ for drama.

As such, in order to measure de Sitter-induced drama by the mechanism we have outlined in this paper, our detector must be super-massive. The scale (\ref{mtwoconddS}) is of order 
\be\label{bigmass}
g_{\text{eff}}^2 10^{60}M_\text{P}\sim g_{\text{eff}}^2 10^{20} M_{\text{sun}},
\ee
where $g_{\text{eff}}=(M_\text{P}\sqrt{\alpha'})^{-1}\sim g_\text{s}/\sqrt{\text{Vol}(M_{\text{internal}})}$ is the ratio of the string and four-dimensional Planck mass scales (here $g_{\text{s}}$ is the fundamental string coupling and $\text{Vol}(M_{\text{internal}})$ is the volume of the extra dimensions in string theory). 
For $g_{\text{eff}}\sim 1/10$, the mass scale (\ref{mtwoconddS}) is higher than super-cluster masses.  However, for smaller $g_{\text{eff}}$, which can arise from larger internal dimensions, the mass scale (\ref{bigmass}) is reduced accordingly.
Cosmic strings could naturally exhibit this scale of mass, but since these have not been detected this is no contraint on our mechanism as it stands.    
 
Even in the more exotic scenarios where detectors satisfying (\ref{mtwoconddS}) arise, the timescale required by (\ref{tHlargedS}) is greater than a Hubble time $H^{-1}\sim L$.  The late universe is only now entering an accelerating phase, and thus this condition is only marginally satisfied at best.  It is tempting to entertain this as being potentially related to the `why now' problem, but for now we simply note this is not a constraint on the mechanism as applied to the observed universe, and we leave further applications to later work.

%\subsection{Late universe acceleration}

%* $\Delta t$ marginal 

%\subsection{Early universe inflation}

The mechanism is even less constrained in the early universe.  In inflationary theory, there is no reason for infallers to exist to produce or detect drama, as all measurements are consistent with an effective initial condition described by the Bunch-Davies vacuum.  As in the late universe, more exotic scenarios which contain additional matter, such as \cite{trapped}, may in principle be constrained.

To summarize, because of the conditions on the detector mass and timescales required for detectable, de Sitter-induced drama, our mechanism is not significantly constrained by cosmological observables in the `vanilla' scenarios for early and late universe cosmology.  More elaborate theories may be subject to interesting constraints, and we noted a potential connection to the `why now' problem.    

\section{Conclusions, caveats, and future directions}

In this work, we first refined our understanding of the longitudinal spreading effect proposed in the light cone calculations \cite{lennyspreading}.  In particular, we noted that for the case of string-detected string spreading, the $2\to 2$ calculation in \cite{bpst}\ in the regime $s\gg -t\gg 1/\alpha'$ exhibits very precisely a mode cutoff of the kind anticipated in \cite{lennyspreading}.  We found that a factor of $\alpha' t$ suppresses the detectable spreading relative to the original estimate \cite{lennyspreading}\ which was based on the nominal resolution time of the detector.  Perhaps not surprisingly, this indicates that other aspects of the detection process figure into the detected spreading.      

With this understanding, we applied the criterion for detectable longitudinal spreading to a late infalling detector in a black hole background which
also contains early-infalling string matter.  In the near horizon region, the metric is flat and the calculation \cite{lennyspreading}\ can be applied to the early string wavefunction.   As the time separation between the two infallers grows, they develop a large relative boost which compensates for their large separation along the horizon. This led us to a specific range of parameters (a lower bound on the mass of the infaller) for which the detector -- more specifically secondary probes emitted by it -- could detect the early string.  This range of parameters includes a window in which the effect constitutes a breakdown of effective field theory, one that is caused by the large relative boost induced by the long time evolution of trajectories in the weakly curved black hole geometry.

These estimates for the detected spreading assumes that the early-infalling string evolves into a standard flat spacetime single-string state in the near horizon region, one satisfying (\ref{longspread}).  If instead the (mild) curvature of the geometry were to obstruct the longitudinal spreading itself, then the conclusion could be altered.  In that case, it might be interesting to start with the strings satisfying (\ref{longspread}) in the near horizon region, and evolve
them back outside the black hole to see to what initial state this corresponds, if not the original infalling strings envisioned in our setup.

Flat spacetime S-matrix elements in our companion paper \cite{usSmatrix}\ exhibit simple features which indicate longitudinal nonlocality, assuming the standard transverse distribution is as given by (\ref{transpread}).  These processes, however, do not share the following basic feature we have in the black hole problem.   Once they reach the near-horizon flat space region, the problem in flat space is as if the detector and early string materialize in a configuration well separated along the light cone, moving away from each other.  Such an arrangement of source string and detector is consistent with detectable spreading according to the criteria given above, leading to our estimates in this paper.  But it has not been checked explicitly with an S-matrix calculation, although other S-matrix amplitudes do exhibit strong evidence for longitudinal nonlocality.  It would be interesting to analyze flat space 6-point amplitudes in order to incorporate this aspect of our problem.  

It is also interesting to approach this question using AdS/CFT.\footnote{We thank S. Shenker and D. Stanford for preliminary discussions of this.}  Pure vacuum AdS is a setting which is in some sense intermediate between flat spacetime and black holes.  The Rindler horizon in AdS has some properties in common with a black hole \cite{Emparan}, such as a relative boost exponential in the analogue of the Schwarzschild time coordinate for early and late infallers.   In some contexts, such as the non-adiabatic effects explored in \cite{backdraft}, the additional symmetry of Poincar\'e time translation shuts off the effect, but with more generic source and detector combinations in the present work, it may be that the process could proceed at the Rindler horizon.   By transforming between a Schwarzschild (thermal) and Poincar\'e description\footnote{See e.g. \cite{InsightfulD}\cite{ShenkerStanford}\ for some of the relevant formulas.} one can set up a CFT measurement in the thermal system which would in principle be able to make an indirect detection of the bulk process of the late infaller detecting the early string.  The exponential falloff in time of appropriate late-time thermal CFT correlators can put a general constraint on the extent to which early matter can be detected late, depending on the type of correlator involved \cite{ShenkerStanford}.  However, it is worth noting that our analysis is actually consistent with an exponential falloff.  In the flat spacetime scattering process of \cite{bpst}\ which exhibits the explicit mode cutoff (\ref{nmaxcons}), the saddle point analysis (\ref{Tsaddle}) is only valid for momentum transfers that are greater than string scale.  This flat spacetime Regge amplitude is proportional to $\sim s^{\alpha't }$.  If we apply this estimate of the amplitude to a black hole or AdS,  since $s\propto e^{(t_2-t_1)/2r_s}$ (with $r_s$ the AdS radius for the case of the Rindler horizon), this in itself would lead to an exponential suppression at late times.  If this Regge form for the amplitude does apply to the black hole process, the suppression in amplitude in the regime where the saddle point analysis holds clearly renders the effect less `dramatic' than otherwise,\footnote{But see below for other factors that may enhance the effect numerically.} but it remains a distinct effect from EFT processes.  It would be very interesting to understand how to extend the analysis to smaller $\alpha't$, where the amplitude could be larger, and to compare carefully with appropriate AdS/CFT calculations.  We leave this to future work.

That said, at various points we made conservative choices in estimating the effect.  Its dependence on detector mass in particular remains to be fleshed out more precisely.  We found an effect going beyond EFT in the present work without folding in the plethora of final states (the Hagedorn density), and the summed contribution of early matter forming the black hole.  Especially in contemplating the possibility of real observations, these should be considered.  Finally, many other aspects of this problem remain to be fleshed out, such as generalization to Kerr black holes, analysis of the detectability of early string spreading by outgoing Hawking particles, and exploring potential observational windows as well as further theoretical constraints.  For now, we can conclude that taken at face value, the effect computed in \cite{lennyspreading}\ and refined here via \cite{bpst}\ leads to new effects going beyond EFT for late-infalling probes in black hole physics.

\section*{Acknowledgements}
We would like to dedicate this work to Gary Horowitz on the occasion of his 60th birthday. 
We are especially grateful to Steve Giddings for extensive discussions on several facets of this work.
We thank S. Giddings, D. Harlow, J. Maldacena, A. Puhm, and D. Stanford for comments on the manuscript.
We have also benefitted from numerous discussions with colleagues including Ahmed Almheiri, Thomas Bachlechner, Roger Blandford,  Raphael Bousso, Liam Fitzpatrick, David Gross, Daniel Harlow,  Sean Hartnoll, Veronica Hubeny, Luis Lehner, Juan Maldacena, Don Marolf, Liam McAllister, Joe Polchinski, Mukund Rangamani, Steve Shenker, Douglas Stanford, Jamie Sully, Lenny Susskind,  and Gabriele Veneziano. 
%Gabriele, Thomas, Douglas, Steve S., Lenny, Jamie, Joe, Ahmed, Liam M., Daniel, Gross, Juan, Don, Mukund, Liam F. ...
This research was supported in part by the National Science Foundation under Grant No. NSF PHY11-25915. The work of E.S. was supported  in part by the National Science Foundation
under grant PHY-0756174 and NSF PHY11-25915 and by the Department of Energy under
contract DE-AC03-76SF00515. M.D. is also supported by a Stanford Graduate Fellowship and a KITP Graduate Fellowship.   It is a pleasure to acknowledge the Aspen Center for Physics and the KITP for hospitality during the later portions of this work.  

\begingroup\raggedright\begin{thebibliography}{10}

\baselineskip=14.5pt

\bibitem{backdraft}
E. Silverstein,
``Backdraft: String Creation in an Old Schwarzschild Black Hole," (2014) [hep-th/1402.1486].

A. Puhm, F. Rojas, T. Ugajin, work in progress.

\bibitem{lennyspreading}
L. Susskind,
``Strings, black holes and Lorentz contraction,"
Phys. Rev. D {\bf 49}, 6606-6611 (1994) 
[hep-th/9308139].

\bibitem{JoeBHcomp} 
  J.~Polchinski,
  ``String theory and black hole complementarity,''
  In *Los Angeles 1995, Future perspectives in string theory* 417-426
  [hep-th/9507094].
  %%CITATION = HEP-TH/9507094;%%
  %28 citations counted in INSPIRE as of 04 Feb 2015

%\cite{Giddings:2007ie}
\bibitem{Giddingsboost} 

S.~B.~Giddings and M.~Lippert,
  ``The Information paradox and the locality bound,''
  Phys.\ Rev.\ D {\bf 69}, 124019 (2004)
  [hep-th/0402073].
  %%CITATION = HEP-TH/0402073;%%
  %53 citations counted in INSPIRE as of 15 Apr 2015

 S.~B.~Giddings,
  ``Black hole information, unitarity, and nonlocality,''
  Phys.\ Rev.\ D {\bf 74}, 106005 (2006)
  [hep-th/0605196].
  %%CITATION = HEP-TH/0605196;%%
  %75 citations counted in INSPIRE as of 15 Apr 2015

%\cite{Strominger:1996sh}
\bibitem{StromingerVafa} 
  A.~Strominger and C.~Vafa,
  ``Microscopic origin of the Bekenstein-Hawking entropy,''
  Phys.\ Lett.\ B {\bf 379}, 99 (1996)
  [hep-th/9601029].
  %%CITATION = HEP-TH/9601029;%%
  %1913 citations counted in INSPIRE as of 13 Feb 2015
  
   C.~G.~Callan and J.~M.~Maldacena,
  ``D-brane approach to black hole quantum mechanics,''
  Nucl.\ Phys.\ B {\bf 472}, 591 (1996)
  [hep-th/9602043].
  %%CITATION = HEP-TH/9602043;%%
  %588 citations counted in INSPIRE as of 18 Apr 2015
  
\bibitem{BHstringscatt}

D.~Amati, M.~Ciafaloni and G.~Veneziano,
  ``Superstring Collisions at Planckian Energies,''
  Phys.\ Lett.\ B {\bf 197}, 81 (1987).
  %%CITATION = PHLTA,B197,81;%%
  %438 citations counted in INSPIRE as of 31 Mar 2015

 S.~B.~Giddings, D.~J.~Gross and A.~Maharana,
  ``Gravitational effects in ultrahigh-energy string scattering,''
  Phys.\ Rev.\ D {\bf 77}, 046001 (2008)
  [arXiv:0705.1816 [hep-th]].
  %%CITATION = ARXIV:0705.1816;%%
  %52 citations counted in INSPIRE as of 31 Mar 2015    
  
  \bibitem{firewalls}
A. Almheiri, D. Marolf, J. Polchinski, and J. Sully,
``Black Holes: Complementarity or Firewalls?"
JHEP {\bf 062} 1302 (2013) [hep-th/1207.3123];

A.~Almheiri, D.~Marolf, J.~Polchinski, D.~Stanford and J.~Sully,
  ``An Apologia for Firewalls,''
  JHEP {\bf 1309}, 018 (2013)
  [arXiv:1304.6483 [hep-th]].
  %%CITATION = ARXIV:1304.6483;%%
  %129 citations counted in INSPIRE as of 21 Apr 2015

%\cite{Braunstein:2009my}
%\bibitem{Braunstein:2009my} 
  S.~L.~Braunstein, S.~Pirandola and K.~Życzkowski,
  ``Better Late than Never: Information Retrieval from Black Holes,''
  Phys.\ Rev.\ Lett.\  {\bf 110}, no. 10, 101301 (2013)
  [arXiv:0907.1190 [quant-ph]].
  %%CITATION = ARXIV:0907.1190;%%
  %162 citations counted in INSPIRE as of 31 Mar 2015

\bibitem{sixauthor}
%\cite{Lowe:1995ac}
%\bibitem{Lowe:1995ac} 
  D.~A.~Lowe, J.~Polchinski, L.~Susskind, L.~Thorlacius and J.~Uglum,
  ``Black hole complementarity versus locality,''
  Phys.\ Rev.\ D {\bf 52}, 6997 (1995)
  [hep-th/9506138].
  %%CITATION = HEP-TH/9506138;%%
  %144 citations counted in INSPIRE as of 31 Mar 2015

\bibitem{NVNL}

S.~B.~Giddings,
  ``Nonviolent nonlocality,''
  Phys.\ Rev.\ D {\bf 88}, 064023 (2013)
  [arXiv:1211.7070 [hep-th]].
  %%CITATION = ARXIV:1211.7070;%%
  %38 citations counted in INSPIRE as of 15 Apr 2015

S.~B.~Giddings,
  ``Nonviolent information transfer from black holes: A field theory parametrization,''
  Phys.\ Rev.\ D {\bf 88}, no. 2, 024018 (2013)
  [arXiv:1302.2613 [hep-th]].
  %%CITATION = ARXIV:1302.2613;%%
  %26 citations counted in INSPIRE as of 09 Feb 2015
  %\cite{Giddings:2014ova}
\bibitem{SteveObservations} 
  S.~B.~Giddings,
  ``Possible observational windows for quantum effects from black holes,''
  Phys.\ Rev.\ D {\bf 90}, no. 12, 124033 (2014)
  [arXiv:1406.7001 [hep-th]].
  %%CITATION = ARXIV:1406.7001;%%
  %3 citations counted in INSPIRE as of 09 Feb 2015
  \bibitem{stringsize}
M. Karliner, I. R. Klebanov, and L. Susskind,
``Size and Shape of Strings," 
Int. J. Mod. Phys. {\bf A3} 1981 (1988). 
 \bibitem{lightcone}
 S. Mandelstam,
 ``Interacting String Picture of Dual Resonance Models,"
Nucl. Phys. B {\bf 64} 205-235 (1973).
\bibitem{usSmatrix}
M. Dodelson and E. Silverstein, ``Longitudinal nonlocality in the string S-matrix," to appear. 
\bibitem{lennykogut}
J. Kogut and L. Susskind,
``The parton picture of elementary particles,"
Phys. Rept. {\bf 8}, 75-172 (1973).
\bibitem{bpst}
R. C. Brower, J. Polchinski, M. J. Strassler, C. Tan,
``The Pomeron and gauge/string duality,"
JHEP {\bf{0712}} 005 (2007) [hep-th/0603115].
\bibitem{grossmende}
D. J. Gross and P. F. Mende,
``The High-Energy Behavior of String Scattering Amplitudes,"
Phys. Lett. B. {\bf 197} 129 (1987).
\bibitem{gribov}
V. N. Gribov,
``Space-time description of hadron interactions at high-energies," 
(1973) [hep-ph/0006158].
\bibitem{sup}
T. Yoneya, 
``String Theory and Space-Time Uncertainty Principle,"
 Prog. Theor. Phys. {\bf 103} 1081-1125 (2000) [hep-th/0004074].
       \bibitem{garyjoe}
      G. T. Horowitz and J. Polchinski,
      ``Self Gravitating Fundamental Strings,"
      Phys. Rev. D {\bf 57} (1998) 2557-2563,
       [hep-th/9707170].
       
       %\cite{Polchinski:1995ta}
\bibitem{Kerraccel}
%\cite{Banados:2009pr}
%\bibitem{Banados:2009pr} 
  M.~Banados, J.~Silk and S.~M.~West,
  ``Kerr Black Holes as Particle Accelerators to Arbitrarily High Energy,''
  Phys.\ Rev.\ Lett.\  {\bf 103}, 111102 (2009)
  [arXiv:0909.0169 [hep-ph]].
  %%CITATION = ARXIV:0909.0169;%%
  %136 citations counted in INSPIRE as of 31 Mar 2015  
\bibitem{preAMPS}

S. L. Braunstein, "Black hole entropy as entropy of
entanglement, or it's curtains for the equivalence principle,"
[arXiv:0907.1190v1 [quant-ph]], published as S. L. Braunstein, S.
Pirandola and K. Zyczkowski, "Better Late than Never: Information
Retrieval from Black Holes," Physical Review Letters 110, 101301
(2013)
%%CITATION = ARXIV:0907.1190;%%
  %79 citations counted in INSPIRE as of 20 Feb 2014

S.~B.~Giddings,
  ``Models for unitary black hole disintegration,''
  Phys.\ Rev.\ D {\bf 85}, 044038 (2012)
  [arXiv:1108.2015 [hep-th]].
  %%CITATION = ARXIV:1108.2015;%%
  %26 citations counted in INSPIRE as of 20 Feb 2014

R.~Haag, H.~Narnhofer and U.~Stein,
  ``On Quantum Field Theory in Gravitational Background,''
  Commun.\ Math.\ Phys.\  {\bf 94}, 219 (1984).
  %%CITATION = CMPHA,94,219;%%
  %102 citations counted in INSPIRE as of 20 Feb 2014

G.~'t Hooft,
  ``On the Quantum Structure of a Black Hole,''
  Nucl.\ Phys.\ B {\bf 256}, 727 (1985).
  %%CITATION = NUPHA,B256,727;%%
  %746 citations counted in INSPIRE as of 20 Feb 2014

L.~Bombelli, R.~K.~Koul, J.~Lee and R.~D.~Sorkin,
  ``A Quantum Source of Entropy for Black Holes,''
  Phys.\ Rev.\ D {\bf 34}, 373 (1986).
  %%CITATION = PHRVA,D34,373;%%
  %516 citations counted in INSPIRE as of 20 Feb 2014

R.~Sorkin,
  ``A Simple Derivation of Stimulated Emission by Black Holes,''
  Class.\ Quant.\ Grav.\  {\bf 4}, L149 (1987).
  %%CITATION = CQGRD,4,L149;%%
  %6 citations counted in INSPIRE as of 20 Feb 2014

K.~Fredenhagen and R.~Haag,
  ``On the Derivation of Hawking Radiation Associated With the Formation of a Black Hole,''
  Commun.\ Math.\ Phys.\  {\bf 127}, 273 (1990).
  %%CITATION = CMPHA,127,273;%%
  %84 citations counted in INSPIRE as of 20 Feb 2014

M.~Srednicki,
  ``Entropy and area,''
  Phys.\ Rev.\ Lett.\  {\bf 71}, 666 (1993)
  [hep-th/9303048].
  %%CITATION = HEP-TH/9303048;%%
  %523 citations counted in INSPIRE as of 20 Feb 2014

T.~Jacobson,
  ``Introduction to quantum fields in curved space-time and the Hawking effect,''
  gr-qc/0308048.
  %%CITATION = GR-QC/0308048;%%
  %73 citations counted in INSPIRE as of 20 Feb 2014

S.~D.~Mathur,
  ``The Information paradox: A Pedagogical introduction,''
  Class.\ Quant.\ Grav.\  {\bf 26}, 224001 (2009)
  [arXiv:0909.1038 [hep-th]].
  %%CITATION = ARXIV:0909.1038;%%
  %113 citations counted in INSPIRE as of 10 Jan 2014

S.~G.~Avery,
  ``Qubit Models of Black Hole Evaporation,''
  JHEP {\bf 1301}, 176 (2013)
  [arXiv:1109.2911 [hep-th]].
  %%CITATION = ARXIV:1109.2911;%%
  %17 citations counted in INSPIRE as of 20 Feb 2014

\bibitem{hawking}
S.~W.~Hawking,
  ``Particle Creation by Black Holes,''
  Commun.\ Math.\ Phys.\  {\bf 43}, 199 (1975)
  [Erratum-ibid.\  {\bf 46}, 206 (1976)].
  %%CITATION = CMPHA,43,199;%%
  %4772 citations counted in INSPIRE as of 09 Jan 2014

\bibitem{scrambling}
Y. Sekino and L. Susskind, 
``Fast Scramblers,"
JHEP {\bf{0810}} 065 (2008) [hep-th/0808.2096].

\bibitem{Pagetime}

D.~N.~Page,
  ``Average entropy of a subsystem,''
  Phys.\ Rev.\ Lett.\  {\bf 71}, 1291 (1993)
  [gr-qc/9305007].
  %%CITATION = GR-QC/9305007;%%
  %137 citations counted in INSPIRE as of 30 Dec 2013

D.~N.~Page,
  ``Black hole information,''
  hep-th/9305040.
  %%CITATION = HEP-TH/9305040;%%
  %102 citations counted in INSPIRE as of 30 Dec 2013

  %\cite{Shenker:2014cwa}
\bibitem{ShenkerStanford} 
  S.~H.~Shenker and D.~Stanford,
  ``Stringy effects in scrambling,''
  arXiv:1412.6087 [hep-th].
  %%CITATION = ARXIV:1412.6087;%%
  %2 citations counted in INSPIRE as of 13 Feb 2015
  
  J.~Maldacena, S.~H.~Shenker and D.~Stanford,
  ``A bound on chaos,''
  arXiv:1503.01409 [hep-th].
  %%CITATION = ARXIV:1503.01409;%%
  %3 citations counted in INSPIRE as of 31 Mar 2015

%\cite{Krichbaum:2006hi}
\bibitem{EHT} 
  T.~P.~Krichbaum, D.~A.~Graham, A.~Witzel, J.~A.~Zensus, A.~Greve, M.~Grewing, M.~Bremer and S.~Doeleman {\it et al.},
  ``Towards the Event Horizon: High Resolution VLBI Imaging of Nuclei of Active Galaxies,''
  astro-ph/0607077.
  %%CITATION = ASTRO-PH/0607077;%%
  %1 citations counted in INSPIRE as of 13 Feb 2015
  
  %\cite{Genzel:2014oya}
\bibitem{Solvay} 
  R.~Genzel,
  ``Massive Black Holes: Evidence, Demographics and Cosmic Evolution,''
  arXiv:1410.8717 [astro-ph.GA];
  %%CITATION = ARXIV:1410.8717;%%
  %1 citations counted in INSPIRE as of 13 Feb 2015
  
  %\cite{Spergel:1997vq}
\bibitem{SpergelZaldarriaga} 
  D.~N.~Spergel and M.~Zaldarriaga,
  ``CMB polarization as a direct test of inflation,''
  Phys.\ Rev.\ Lett.\  {\bf 79}, 2180 (1997)
  [astro-ph/9705182].
  %%CITATION = ASTRO-PH/9705182;%%
  %86 citations counted in INSPIRE as of 15 Feb 2015

%\cite{Adam:2015rua}
\bibitem{Planck} 
  R.~Adam {\it et al.}  [Planck Collaboration],
  ``Planck 2015 results. I. Overview of products and scientific results,''
  arXiv:1502.01582 [astro-ph.CO].
  %%CITATION = ARXIV:1502.01582;%%  
%\cite{Spradlin:2001pw}
\bibitem{StromingerdS} 
  M.~Spradlin, A.~Strominger and A.~Volovich,
  ``Les Houches lectures on de Sitter space,''
  hep-th/0110007;
  %%CITATION = HEP-TH/0110007;%%
  %288 citations counted in INSPIRE as of 15 Feb 2015  
  
  A.~Zee,
  ``Einstein Gravity in a Nutshell,''
  %%CITATION = INSPIRE-1230427;%%
  %2 citations counted in INSPIRE as of 15 Feb 2015
  
\bibitem{trapped}
D.~Green, B.~Horn, L.~Senatore and E.~Silverstein,
  ``Trapped Inflation,''
  Phys.\ Rev.\ D {\bf 80}, 063533 (2009)
  [arXiv:0902.1006 [hep-th]].
  %%CITATION = ARXIV:0902.1006;%%
  %82 citations counted in INSPIRE as of 15 Feb 2015
  
 \bibitem{Emparan}
 
 %\cite{Emparan:1999gf}
%\bibitem{Emparan:1999gf} 
  R.~Emparan,
  ``AdS / CFT duals of topological black holes and the entropy of zero energy states,''
  JHEP {\bf 9906}, 036 (1999)
  [hep-th/9906040].
  %%CITATION = HEP-TH/9906040;%%
  %111 citations counted in INSPIRE as of 17 Apr 2015
  
  \bibitem{InsightfulD}
  
  G.~Horowitz, A.~Lawrence and E.~Silverstein,
  ``Insightful D-branes,''
  JHEP {\bf 0907}, 057 (2009)
  [arXiv:0904.3922 [hep-th]].
  %%CITATION = ARXIV:0904.3922;%%
  %46 citations counted in INSPIRE as of 17 Apr 2015

\endgroup

\end{document}